\documentclass[twocolumns]{aa}  
\usepackage{amsmath}
\usepackage{graphicx}
\usepackage[usenames]{color}
\usepackage{natbib} 
\bibpunct{(}{)}{;}{a}{}{,} 
\usepackage[breaklinks,colorlinks=true,citecolor=black,linkcolor=black,urlcolor=blue,bookmarks=true]{hyperref}
\usepackage{breakurl}  

\usepackage{txfonts}
\begin{document}
   \title{On recent SFR calibrations and the constant SFR approximation}

   \author{
          M. Cervi{\~n}o$^{1,2,3}$
         \and
          A. Bongiovanni$^{1,2,4}$
          \and
          S. Hidalgo$^{1,2}$
          }

\offprints{M. Cervi{\~n}o, \email{mcs@iaa.es}}
\institute{Instituto de Astrof{\'{\i}}sica de Canarias, c/ v{\'{\i}}a
  L\'actea s/n, 38205 La Laguna, Tenerife, Spain 
\and Departamento de Astrof{\'{\i}}sica, Universidad de La Laguna
(ULL), 38205 La Laguna, Tenerife, Spain 
\and Instituto de Astrof{\'\i}sica de Andaluc{\'\i}a (IAA-CSIC),
Placeta de la Astronom{\'\i}a s/n, E-18008 Granada, Spain 
\and Asociaci\'on ASPID, Apartado de Correos 412, 38200 La Laguna,
Tenerife, Spain }

\date{\today}

 
  \abstract
{}
{Star Formation Rate (SFR) inferences are based in the so-called
  constant SFR approximation, where synthesis models are require to
  provide a calibration; we aims to study the key points of such
  approximation to produce accurate SFR inferences.}
{We use the intrinsic algebra used in synthesis models, and we explore
  how SFR can be inferred from the integrated light without any
  assumption about the underling Star Formation history (SFH).}
{We show that the constant SFR approximation is actually a simplified
  expression of more deeper characteristics of synthesis models: It is
  a characterization of the evolution of single stellar populations
  (SSPs), acting the SSPs as sensitivity curve over different measures
  of the SFH can be obtained.  As results, we find that (1) the best
  age to calibrate SFR indices is the age of the observed system
  (i.e. about 13 Gyr for $z=0$ systems); (2) constant SFR and
  steady-state luminosities are not requirements to calibrate the SFR;
  (3) it is not possible to define a SFR single time scale over which
  the recent SFH is averaged, and we suggest to use typical SFR
  indices (ionizing flux, UV fluxes) together with no typical ones
  (optical/IR fluxes) to correct the SFR from the contribution of the
  old component of the SFH, we show how to use galaxy colors to quote
  age ranges where the recent component of the SFH is stronger/softer
  than the older component.}
{Particular values of SFR calibrations are (almost) not affect by this
  work, but the meaning of what is obtained by SFR inferences does. In
  our framework, results as the correlation of SFR time scales with
  galaxy colors, or the sensitivity of different SFR indices to sort
  and long scale variations in the SFH, fit naturally. In addition,
  the present framework provides a theoretical guide-line to optimize
  the available information from data/numerical experiments to improve
  the accuracy of SFR inferences.}  
\maketitle

%

\section{Introduction}
\label{sec:SFRindic}

The knowledge of the amount of gas transformed into stars as a
function of time, so called the star formation history, (SFH,
$\psi(t)$), or at least the amount of gas transformed in stars
recently (star formation rate, SFR, $\psi(\mathrm{t_{now}})$, or
$\psi(t)$ averaged over a recent time inteval) is one of the key
points to understand galaxy evolution and how and when the gaseous
mass has been assembled into stars over cosmic times \cite[see][ for a
  recent review]{MD14}.  The question about the {\it evolution of the
  gas and stars in galaxies} is a broad research area which is
described in a formal way (evolutionary population synthesis models)
in seminal papers as the one by B.~\cite{Tins80}.  The formalism
presented in the 80's had remained practically identical up to present
days, being developments related with the use of observations to
restrict the theoretical parameter space, or to use the models as a
tool to infer physical parameters from observed quantities, as it is
the case of SFR inferences.

The methodology used in recent SFH inferences is driven by
observational trends of galaxy colors \cite[see][]{Ken98}, being
evolutionary synthesis models used to calibrate the relation between a
suitable observed integrated luminosity ${\cal L}_\mathrm{ind}$ and
the recent SFH associated to such luminosity, ${\cal
  {SFR}}_\mathrm{ind}$.

Using the so-called constant SFR approximation \citep{Ken98}, it is
assumed a constant SFH up to an age $\mathrm{t_{test}}$, so suitable
luminosities are these ones which reach a quasi steady-state value
$\ell_{\mathrm{c}\cal{SFR},\mathrm{ind}}^{\mathrm{asymp}}$ after some
age $\mathrm{t_{ind}}$ lower than $\mathrm{t_{test}}$. Provided that
the age $\mathrm{t_{ind}}$ is low enough, the term ``recent" can be
applied.

This situation can be described in general, independently of the final
$\mathrm{t_{ind}}$ value, by the condition

\begin{equation}
\ell_{\mathrm{c}\cal{SFR},\mathrm{ind}}(\mathrm{t_{ind}}) \simeq
\ell_{\mathrm{c}\cal{SFR},\mathrm{ind}}(t) \,\,\,\forall t \in
    [\mathrm{t_{ind}},\mathrm{t_{test}}],
\label{eq:cdef1conts}
\end{equation}

\noindent although such mathematical refinement is usually not taken
into consideration since an asymptotic behavior can we observed by a
naked-eye inspection by plotting the time evolution of
$\ell_{\mathrm{c}\cal{SFR},\mathrm{ind}}(t)$ produced by the models,
or by inspection of the numerical values given by the corresponding
tables. As final result, the value
$\ell_{\mathrm{c}\cal{SFR},\mathrm{ind}}(\mathrm{t_{test}})$ is used
as the asymptotic luminosity
$\ell_{\mathrm{c}\cal{SFR},\mathrm{ind}}^{\mathrm{asymp}}$, since,
actually, $\mathrm{t_{ind}}$ is not required to be computed explicitly
(in addition it avoids further complications about to give a
quantitative meaning to the symbol ``$\simeq$" used in
Eq.~\ref{eq:cdef1conts}; but see below).

Given that $\ell_{\mathrm{c}\cal{SFR},\mathrm{ind}}^{\mathrm{asymp}}$
is obtained under a constant SFH assumption and normalized to a
suitable SFR value (typically 1 M$_\odot$/yr), we can obtain the
associated SFR, $\mathrm{c}\cal{SFR}_\mathrm{ind}$, from the observed
integrated luminosity ${\cal L}_\mathrm{ind}$ as,

\begin{equation}
{\mathrm{c}\cal{SFR}}_\mathrm{ind} = {\cal L}_\mathrm{ind} \times
C_\mathrm{ind},
\label{eq:SFRindex}
\end{equation}

\noindent being 

\begin{equation}
C_\mathrm{ind} = \frac{1}{
  \ell_{\mathrm{c}\cal{SFR},\mathrm{ind}}^{\mathrm{asymp}} } =
\frac{1}{ \ell_{\mathrm{c}\cal{SFR},\mathrm{ind}}(\mathrm{t_{test}})}
\label{eq:cdef1}
\end{equation}

In this methodology the main relevant parameter is
$\mathrm{t_{test}}$, which combines (1) our confidence that an
asymptotic value had been actually reached at $\mathrm{t_{test}}$, and
(2) our believes about how a constant SFH is a valid approximation. As
a reasonable compromise, $\mathrm{t_{test}}$ is chosen to conciliate
both expectations, being a typical value $\mathrm{t_{test}} = 100
\mathrm{Myr}$ \cite[e.g.][]{Ken98,Muretal11}. This choice of
$\mathrm{t_{test}}$ can be justified as (a) the typical life-times of
massive stars which produce each particular SFR proxy ${\cal
  L}_\mathrm{ind}$, so we expect that $\mathrm{t_{test}} \sim
\mathrm{t_{ind}}$, and (b) an age range large enough to include a
large amount of burst-like star forming events formed at different
ages (which at global level approaches to a constant SFH), so the
obtained SFR represents an average of the SFH in the last
$\mathrm{t_{test}}$ time interval. Implicitly is assumed that stars
older that $\mathrm{t_{test}}$ almost do not contribute to ${\cal
  L}_\mathrm{ind}$, and since population colors reders with age, it is
expected that the bluer the galaxy the better the inference about the
real SFR \citep{Ken98}. The proposed calibration was originally
establish for the disk component or irregular type galaxies, so
implicitly a correction from bulge component of the galaxy (the old
componet contribution) should be required; however, the calibration
has been applied extensively to any kind of galaxy (e.g. galaxy
surveys) where such decomposition is not possible.

Maybe the principal characteristic of this approach is that, besides
its simplicity and intrinsic assumptions, it provides a reasonably
good SFR inferences in more wider situations than the ones implicit in
the formulation, including situations where the recent SFH is clearly
not constant \citep[e.g.][ where graphical illustrative examples can
  be found]{BBP14}.  Even more, it looks surprising that, although any
star, whatever its age and initial mass, emits in the whole wavelength
range, the overall contribution of old stars in the system looks to
have almost no impact in current proxies of the SFR except in the
cases of low SFR, \cite[see discussion about in Sect. 5.4
  of][]{Con13}, or for SFR indices related with dust emission at
infrared wavelengths \cite[see discusion in][as an example]{HBI03} So,
although not perfect, the methodology includes the main (and
principal) ingredients required to estimate a SFR, and, depending the
proxy, it would refer to the instantaneous SFR
($\psi(\mathrm{t_{now}}))$ or the averaged SFH over a recent time
interval.

Recent observational developments (in both sensitivity and spatial
resolution) have lead to the requirement of improving calibrations,
and a lot effort has been done in this direction covering different
aspects of the problem \cite[see][~as reviews in the
  subject]{Cal13,KE12,MD14}. As examples about improvements of
$\mathrm{t_{test}}$ let us mention \cite{BBP14}, who propose to use
$\mathrm{t_{test}} = 1~\mathrm{Gyr}$ to produce more accurate results
of $C_\mathrm{ind}$ for {\sc galex}/FUV, NUV and {\sc sdss}/$u$
indices due to the tiny, but non null, contribution to the integrated
luminosities due to stars with ages between 100 Myr to 1 Gyr; or
\cite{Johetal13}, who use $\mathrm{t_{test}} = 10~\mathrm{Gyr}$ to
better match the SFR properties of a sample of (primary) dwarf
galaxies where the SFH had been obtained from CMD analysis. Related
with it, there are the efforts to characterize the time-scale over the
SFR is measured, as the use of a luminosity-weighted effective age
\cite[][]{Buzz02b,BBP14}, or to evaluate the accuracy of
$C_\mathrm{ind}$ by computing explicitly characteristics time scales
$\mathrm{t_{ind, x\%}}$ where some percentage $x$ of the integrated
light is produced, and comparing it with similar time scales obtained
from SFH inferred using different methodologies (as CDMs analysis or
SED fitting); see discussions in \cite{Leretal12},
\cite{Haoetal11,Cal13,Johetal13} or \cite{Simetal14} as examples.

However, in most cases, the improvements of the calibration deals with
the computation of evolutionary synthesis models using more or less
sophisticated SFH to obtain the final (numerical) values and to
compare them with the numerical values obtained under the constant SFR
(and $\mathrm{t_{test}}$) hypothesis; that is, the focus is placed in
the variations of $C_\mathrm{ind}$ (or characteristics time scales) in
different situations. And, afterwards, and despite variations due to
fluctuations of the recent SFH at short time-scales
\cite[e.g.][]{OFMH10}, the constant SFR approximation looks to be a
quite good job (modulus the choice of $\mathrm{t_{test}}$). So, In
this work we ask ourselves the following questions: Is there any
theoretical argument to define an optimal value of
$\mathrm{t_{test}}$?  Is a constant SFR requirement fundamental for
the calibration? If yes, why do the calibrations work even for SFHs
that are raelly not constant?; if not, what is the physical meaning of
the inferred value ${\mathrm{c}\cal{SFR}}_\mathrm{ind}$?

To answer such questions we require to understand how the SFH is
implemented in synthesis codes in first instance; which is done in
Sect.\ref{sec:SBR}. Secondly, we must understand what a synthesis code
will provide independently of any specific choice of the SFH, and
define the problem of SFR inferences using the algebra associated to
synthesis models. To do so, we use first reasonable analytical
approximations which provides hints and guidelines about different
aspects of SFR calibrations/inferences \cite[we recommend the woks
  of][which illustrates nicely this
  approach]{Tins80,Buzz02,Buzz02b,Buzz05}; before to compute the
calibrations explicitly.  This proccess is shown in
Sect.~\ref{sec:SFR}. In Sect.~\ref{sec:disc}Ê~we apply the analysis
about SFRs obtained in this work to corroborate and extend some
results about SFR calibrations obtained recently. Our conclusions are
presented in Sect.~\ref{sec:conclu}.  In companion papers we will
investigate the explicitly the sensitivity of SFR calibrations to the
different choices of synthesis models (which results are briefly
summarized in Sect.~\ref{sec:SFR}), and to how the overall SFH would
affect (recent) SFR inferences.  The overall idea developed in this
paper implies to dismount some of the (unnecessary) assumptions about
SFR inferences, so each section is written in schematic fashion.

\section{SFH implementation in synthesis models}
\label{sec:SBR}

{\bf 1.} Evolutionary synthesis models are designed to describe the
spectrophotometric evolution ${\cal L}_\lambda(t)$ of stellar
ensembles (independently that other quantities are also obtained) for
a given initial conditions.  In our context, the initial conditions
are some recipe providing how many stars of different initial masses
had been formed at different time, (that is the stellar birth rate
${\cal{B}}(m,t)$), and the relation between luminosity at a given
band/wavelength of an star given its initial mass and its evolutionary
age $t_*$, $\ell_\lambda(m,t_*)$ \footnote{Actually it should read
  $\ell_\lambda(m,t_*,Z,\Omega)$, being $Z$ the initial metallicity of
  the star, and $\Omega$ its rotational velocity, which implies to
  include the corresponding parameters in the stellar birth-rate. In
  addition, it can be also considered interactions between stars
  (i.e. binary interactions), which depend on additional parameters
  that, again, must be included in the stellar birth-rate. Along this
  work we neglect all such additional parameters.}.

{\bf 2.} Typically is assumed that ${\cal{B}}(m,t)$ can be decomposed
in two independent functions, the one giving the frequency
distribution of the initial masses of stars that would be formed
whatever the age (it is, the initial mass function , IMF, $\phi(m)$)
and other giving the the amount of stars formed at each time (it is,
the star formation history, SFH, $\psi(t)$).  The mass range where the
stellar birth rate (hence the IMF) is defined must cover all
physically posible stars formed $[\mathrm{m_{low}},\mathrm{m_{up}}]$
and it is imposed by stellar physics. The time range where the stellar
birth rate (hence the SFH) is defined must include all the posible
ages when a star of any mass would had been formed in the ensemble,
so, in practical terms it covers from the time $\mathrm{t_{ini}}$ when
the first star is formed in the observed system, to the (rest-frame)
time where the observation is done $\mathrm{t_{now}}$. In the case of
galaxies and stellar ensembles inside galaxies, the value of
$\mathrm{t_{ini}}$ is given by cosmological studies as far as we
accept that there is an epoch of galaxy formation, and that any
stellar ensemble inside a galaxy would contain a relic contribution of
the first formed stars (a quite plausible assumption which depends on
the movements/redistribution of stars formed at different times due to
galactic dynamics). Finally, the value of $\mathrm{t_{now}}$ is
imposed by the observation the redshift of the source and the choice
of a cosmological model.

Being ${\cal{B}}(m,t)$ defined only in a time interval, we can define
the age of the ensemble as the time interval since the first star has
been formed up to the rest-frame present time, i.e. $\mathrm{t_{age}}
= \mathrm{t_{now}} -\mathrm{t_{ini}}$, encoding in it all the
cosmological considerations; so ${\cal{B}}(m,t)$ is defined as
$[0,\mathrm{t_{age}}]$ being $t$ the proper age of the global system.
Assuming that both ${\cal{B}}(m,t)$ and $\ell_\lambda(m,t_*)$ are well
comported and integrable functions, and taking into account that a
star born at a time $t$ has an estellar age $t_* = \mathrm{t_{age}} -
t$, the resulting luminosity of the ensemble
${\cal{L}}_\lambda(\mathrm{t_{age}})$ at any $\mathrm{t_{age}}$ value
is obtained as:

\begin{eqnarray}
{\cal{L}}_\lambda(\mathrm{t_{age}}) & = & \int_0^{\mathrm{t_{age}}}
\int_{\mathrm{m_{low}}}^{\mathrm{m_{up}}}
\ell_\lambda(m,\mathrm{t_{age}} - t)\, {\cal{B}}(m,t)\,\,
\mathrm{d}m\, \mathrm{d}t \,\, \nonumber \\ &=&
\int_0^\mathrm{t_{age}} \left[
  \int_{\mathrm{m_{low}}}^{\mathrm{m_{up}}}
  \,\ell_\lambda(m,\mathrm{t_{age}} - t)\, \phi(m)
  \mathrm{d}m\,\right] \psi(t)\, \mathrm{d}t \,\, \nonumber \\ &=&
\int_0^\mathrm{t_{age}} \ell_{\lambda,\mathrm{IMF}}(\mathrm{t_{age}} -
t)\, \psi(t)\, \mathrm{d}t
\label{eq:Ltot}
\end{eqnarray}

\noindent where the term $\ell_{\lambda,\mathrm{IMF}}(\mathrm{t_{age}}
- t) = \ell_{\lambda,\mathrm{IMF}}(t_*) $ refers to the integrated
luminosity when only stars with the same stellar age $t_*$ are
considered. Since such situation can be also described as the
resulting luminosity when the SFH is described as a Dirac's delta
distribution, this quantity is usually referred as the integrated
luminosity of a single age (single metallicity) stellar population, or
SSP. Although we use the term SSP along the work, since commonly used
in the literature, we keep the notation
$\ell_{\lambda,\mathrm{IMF}}(t_*)$, which explicitly shows that such
result actually does not include information about the SFH, neither it
represents an integrated luminosity but just an useful mathematical
entity which only contains information about the stellar evolution
(and $\phi(m)$), which is always well defined.

{\bf 3.} A simple inspection of Eq.~\ref{eq:Ltot} shows that
${\cal{L}}_\lambda(\mathrm{t_{age}})$, which is the only observable
quantity, is always evaluated over the complete age range where the
SFH $\psi(t)$ and the SSP luminosities
$\ell_{\lambda,\mathrm{IMF}}(t_*)$ are defined. Any stellar population
synthesis computation including all stellar evolutionary phases shows
that $\ell_{\lambda,\mathrm{IMF}}(t_*)$ never reach a zero value, so

\begin{quote}
the most plausible $\mathrm{t_{test}}$ to be used to calibrate recent
SFR indices is the age of the system $\mathrm{t_{age}}$ (which has a
value around 13 Gyr in the local Universe), since it is the intrinsic
time provided by the observable luminosity.
\end{quote}

We cannot escape from this result: Whatever the observable luminosity,
it includes the contribution of stars covering all possible range of
stellar ages $t_*$ from 0 (just born stars at $\mathrm{t_{now}}$) to
$\mathrm{t_{age}}$ (the first formed stars in the system that are
still alive). There is no way to discriminate the contribution of
stars with different ages without knowing the whole SFH, or
equivalently, we cannot calibrate a SFR by constraint the SFH to our
concept of "recent" encoded in a $\mathrm{t_{test}}$ value. The result
would be shocking for some readers, since following literally the
methodology to calibrate the SFR, it would imply to assume a constant
SFH all over the life-time of the galaxy; a result that hardly
conciliates with our current understanding of galaxy evolution. And
ever more shocking taking into account that the calibrations used in
the literature, although assume a $\mathrm{t_{test}}$ much lower than
$\mathrm{t_{age}}$, produce in average a quite good job.

{\bf 4.}  The solution to such apparent muddle is to change the
perspective about the role of the SFH in the calibration of SFR
indices: the approximation used to calibrate SFR indices does not deal
with any particular SFH, but with $\ell_{\lambda,\mathrm{IMF}}(t_*)$;
a constant SFH assumption is equivalent to use no information at all
about the SFH.  Would be ${\cal{L}}_\lambda(\mathrm{t_{age}})$
produced by, and only by, stars with ages $t_*$ equal or lower than
$\mathrm{t_{ind}}$, then, whatever the functional form of $\psi(t)$,
the associated integrated luminosity is the result of the SFH
restricted to the time interval
$[\mathrm{t_{now}}-\mathrm{t_{ind}},\mathrm{t_{now}}]$. Even more,
such ${\cal{L}}_\lambda(\mathrm{t_{age}})$ reach a steady state for
any $\mathrm{t_{age}} > \mathrm{t_{ind}}$, and a
$\ell_{\lambda,\mathrm{IMF}}(t_*)$-weighted averaged SFH over the last
$\mathrm{t_{ind}}$ age range (i.e. an SFR) can be obtained.  So,
although under such conditions, we can translate the situation to
consider the SFH only defined up to $\mathrm{t_{ind}}$ and we will
obtain the same result, it is the characteristics of the chosen
luminosity (i.e. of $\ell_{\lambda,\mathrm{IMF}}(t_*)$) what allows to
obtain SFR inferences, not the choice of any particular SFH. As
result, the SFR calibration is actually a characterization of the
evolution of SSP, $\ell_{\lambda,\mathrm{IMF}}(t_*)$, instead a
question about the choice of a $\mathrm{t_{test}}$ value and $\psi(t)$
functional forms typically addressed in the literature.  Let us
exploit this idea in the following section.

\section{The SFR calibration as a characterization of the evolution of SSP luminosity, $\ell_{\lambda,\mathrm{IMF}}(t_*)$, instead a constant SFR hypothesis}
\label{sec:SFR}

To fully exploit the statement and implications quoted in the previous
section it is required a step by step process. In the following, let
us use Eq.~\ref{eq:Ltot} with different (hypothetical and realistic)
$\ell_{\lambda,\mathrm{IMF}}(t_*)$ functional forms to obtain results
about SFR inferences. We stress that all along this section no
hypothesis about the SFH is required.

\subsection{SSP luminosity evolving as a hat function}

{\bf 1.} As a first simple example, let us assume that the SSP
luminosity $\ell_{\lambda,\mathrm{IMF}}(t_*)$ evolves as is a
hat-function with a constant value $\ell_{\lambda,\mathrm{cte}}$ in a
given time range $[t_{*,\mathrm{begin}},t_{*,\mathrm{end}}]$, hence
covering a time interval $\Delta t = t_{*,\mathrm{end}} -
t_{*,\mathrm{begin}}$, and zero otherwise. Trivially Eq.~\ref{eq:Ltot}
is only defined in the time interval
$[\mathrm{t_{age}}-t_{*,\mathrm{end}},\mathrm{t_{age}}-t_{*,\mathrm{begin}}]$
and, after some trivial operations,

\begin{equation}
\left<{\cal{SFR}}\right>_{t_{*,\mathrm{end}},{\Delta t}} = \frac{
  \int_{\mathrm{t_{age}} - t_{*,\mathrm{end}}
  }^{\mathrm{t_{age}}-t_{*,\mathrm{begin}}} \psi(t) \mathrm{d}t
}{\Delta t} =\frac{
  {\cal{L}}_\lambda(\mathrm{t_{age}})}{\ell_{\lambda,\mathrm{cte}}
  \times \Delta t},
\label{eq:5}
\end{equation}

\noindent where $\left<{\cal{SFR}}\right>_{t_{*,\mathrm{end}},{\Delta
    t}}$ is exactly the mean value of the SFH in the corresponding
time interval where $\ell_{\lambda,\mathrm{IMF}}(t_*)$ is
defined. Note that, in order to understand such measure, two
quantities are required: the associated time interval and one of the
time boundaries. Trivially, if $t_{*,\mathrm{begin}} = 0$, we have
$\Delta t = {t_{*,\mathrm{end}}}$ and only one parameter is
needed. Let us denote such situation by maning ${t_{*,\mathrm{end}}}$
as ${t_{*,\mathrm{ind}}}$ and
$\left<{\cal{SFR}}\right>_{t_{*,\mathrm{end}},{\Delta t}}$ as
$\left<{\cal{SFR}}\right>_{{\Delta t}}$. In this situation,
$\left<{\cal{SFR}}\right>_{{\Delta t}}$ is an exact measure of the
mean recent SFR in the last $\Delta t = t_{*,\mathrm{ind}}$ time
range.

This measure of the SFR is completely independent of the details of
the SFH functional form, working even for the case of a burst of star
formation where $\psi(t)$ is described as a Dirac's delta function
with intensity $\cal M$: If such event happens in the quoted time
interval, then ${\cal{L}}_\lambda(\mathrm{t_{age}}) = {\cal{M}} \times
\ell_{\lambda,\mathrm{cte}}$, and the mean value of $\psi(t)$ in such
time interval is ${\cal{M}}/\Delta t$.

{\bf 2.} Although a hat-function would be seen as an unrealistic case,
this kind of distribution is similar to the description of how recent
SFR is inferred from Young Stellar Objects (YSO) number counts
$N_\mathrm{YSO}$, which is typically used to introduce SFR inferences
\cite[e.g.][]{Cal13,KE12}. In that case it is only required a time
scale $\tau_\mathrm{YSO}$ where a YSO would be observed (which is
given by the physics of star formation, which has a value around 2
Myr; see \citealt{McKO07} or \citealt{Evanetal09} as examples). So the
SFR inferred from the observation of $N_\mathrm{YSO}$ YSO in units of
number of stars formed by unit time is:

\begin{eqnarray}
\left< {\cal{SFR}}\right>_{\tau_\mathrm{YSO}} &  = & \frac{N_\mathrm{YSO}}{\tau_{\mathrm{YSO}}}, 
\label{eq:SFR_YSOa}
\end{eqnarray}

Implicitly we are neglecting the information that the luminosity of
each YSO would provide about when such object is formed, which is
equivalent to assume {\it de facto} that a hat function defined in the
time interval $[0,\tau_{\mathrm{YSO}}]$. Hence, independently of
posible variations of $\psi(t)$ in such time interval, a correct
average $\left< {\cal{SFR}}\right>_{\tau_\mathrm{YSO}}$ is
obtained\footnote{Actually the time dependence of the luminosity is
  used in works dealing with the star formation process itself where
  different classes of YSO are considered; see
  \cite{Ladetal13,RZetal15} as examples.}.

{\bf 3.} Although we know that no SSP luminosity
$\ell_{\lambda,\mathrm{IMF}}(t_*)$ evolves as a hat function, the hat
function case shows that we cannot obtain $\psi(\mathrm{t_{now}})$
from observations, but, at best, an average value over a time interval
~$\left< {\cal{SFR}}\right>_{\Delta t}$. We can extend the concept of
average the SFR over a time interval, to the concept of obtain a
weighted mean of $\psi(t)$ over any arbitrary function
$\varphi_\lambda(t)$. The only requirement is that such function is
normalized over the time interval $\psi(t)$ is defined
(i.e. $\mathrm{t_{age}}$). In the context of this paper, we can define
the weight function $\varphi_\lambda(t)$ as:

\begin{equation}
\varphi_\lambda(t) =
\frac{\ell_{\lambda,\mathrm{IMF}}(\mathrm{t_{age}} -
  t)}{\int_0^{\mathrm{t_{age}}}
  \ell_{\lambda,\mathrm{IMF}}(\mathrm{t_{age}} - t)\, \mathrm{d}t} =
\frac{\ell_{\lambda,\mathrm{IMF}}(\mathrm{t_{age}} -
  t)}{\int_0^{\mathrm{t_{age}}} \ell_{\lambda,\mathrm{IMF}}(t_*)\,
  \mathrm{d}t_*}.
\label{eq:7}
\end{equation}

So the SFH $\varphi_\lambda(t)$-weighted mean,
$\left<{\cal{SFR}}\right>_\lambda$, is:

\begin{eqnarray}
\left<{\cal{SFR}}\right>_{\lambda} &=& \int_0^{\mathrm{t_{age}}} \,
\psi(t) \, \varphi_\lambda(t) dt = \nonumber \\ &=&
\frac{\int_0^{\mathrm{t_{age}}} \psi(t)\,
  \ell_{\lambda,\mathrm{IMF}}(\mathrm{t_{age}} - t)\,
  \mathrm{d}t}{{\int_0^{\mathrm{t_{age}}}
    \ell_{\lambda,\mathrm{IMF}}(t_*)\, \mathrm{d}t_*}} \nonumber \\ &
=&
\frac{{\cal{L}}_\lambda(\mathrm{t_{age}})}{\int_0^{\mathrm{t_{age}}}
  \ell_{\lambda,\mathrm{IMF}}(t_*)\, \mathrm{d}t_*} = C_{\lambda}
\times {\cal{L}}_\lambda(\mathrm{t_{age}}).
\label{eq:meanSFRok}
\end{eqnarray}

The normalization coefficient of the function $\varphi_\lambda(t)$ is
the inverse of the quantity $C_{\lambda} \equiv C_\mathrm{ind}$ used
in the usual calibrations of the SFR. Of course, such normalization
coefficient can be also interpreted as the luminosity obtained by a
synthesis model under a constant SFH assumption, but actually

\begin{quote} 
a constant SFR assumption is not a requirement to calibrate SFR
indices. It is the evolution of the SSP luminosity (the
$\ell_{\lambda,\mathrm{IMF}}(t_*)$ function) normalized over the
system age, not a hypothesis about the SFH $\psi(t)$, which gives the
meaning to the $\left<{\cal{SFR}}\right>_{\lambda}$ that can be
obtained from observations.
\end{quote}

An alternative interpretation to Eq.~\ref{eq:meanSFRok} is that the
observed luminosity ${\cal{L}}_\lambda(\mathrm{t_{age}})$ is the
result of the SFH $\psi(t)$ once filtered over the evolution of the
luminosity produced by coeval stars $\ell_{\lambda,\mathrm{IMF}}(t_*)$
(defined up to $t_* = \mathrm{t_{age}}$). So, we can obtain direct
information about $\psi(t)$ once the filter is normalized/calibrated,
or, in general grounds, when the zero point of the filter is defined
in a similar way that in photometric studies\footnote{The analogy of
  synthesis models results and photometry is not new and it is quoted
  by \cite{Shore02} Chap. 7, Sect. 3.3; surprising such analogy has
  been poorly explored in the literature, and typically limited to
  restricted $\mathrm{t_{test}}$ values; however, see
  \cite{OFMH10,Leretal12} as counterexamples who, the facto, use
  $\ell_{\lambda,\mathrm{IMF}}(t)$ as SFH-sensitivity curve.}.

{\bf 4.} Previous result is general, so, if we hope that
$\left<{\cal{SFR}}\right>_{\lambda}$ contains only information about
the recent SFH, we require a filter only sensitive to recent
ages. That is, an hypothetical ${\cal{L}}_\lambda(\mathrm{t_{age}})$
which associated $\ell_{\lambda,\mathrm{IMF}}(t_*)$ has a zero value
after some age $t_{*,\mathrm{ind}}$. Such break, if exists, can be
obtained by a direct inspection of $\ell_{\lambda,\mathrm{IMF}}(t_*)$,
but also by the variation over $t_*$ of the integral of
$\ell_{\lambda,\mathrm{IMF}}(t_*)$. Trivially, if it goes to zero
after at some $t_{*,\mathrm{ind}}$ value, then

\begin{equation}
 \int_0^\mathrm{t_{age}} \ell_{\lambda,\mathrm{IMF}}(t_*)
 \mathrm{d}t_* \equiv \int_0^{\mathrm{t_{*,ind}}}
 \ell_{\lambda,\mathrm{IMF}}(t_*) \mathrm{d}t_* \equiv
 \ell^\mathrm{asymp}_{\lambda,\mathrm{c}{\cal{SFR}}} \,\,\, \forall t
 > t_{*,\mathrm{ind}}, \nonumber
\end{equation}

\noindent being $\ell^\mathrm{asymp}_{\lambda,\mathrm{c}{\cal{SFR}}} =
\ell_{\lambda,\mathrm{cte}} \times \Delta t$ for the case of a hat
function. We have keep the symbol
$\ell^\mathrm{asymp}_{\lambda,\mathrm{c}{\cal{SFR}}} $ to stress its
similitude with the calibration constant $C_{\mathrm{ind}}$
(Eqs.~\ref{eq:cdef1},~\ref{eq:5},~\ref{eq:meanSFRok}).

{\bf 5.} The use of the integral over $t_*$ instead a direct
inspection of $\ell_{\lambda,\mathrm{IMF}}(t_*)$ would be seen as an
unnecessary complication. However, the
$\ell_{\lambda,\mathrm{IMF}}(t_*)$ obtained by synthesis codes (or
equivalently, the evolution of SSP models) are not hat-like functions,
neither shows a clear well defined $t_{*,\mathrm{ind}}$ value. Rather
than that, shows that the luminosity declines with $t_*$ more or less
quickly depending on the wavelength. So, if we still aims to obtain a
$\left<{\cal{SFR}}\right>_{\lambda}$ value which can be used as the
actual $\left< {\cal{SFR}}\right>_{\Delta t}$ for some observed
${\cal{L}}_\lambda(\mathrm{t_{age}})$ luminosity, the look for
$\ell_{\lambda,\mathrm{IMF}}(t_*)$ whose integral over time reach a
quasi-state regime is the only approach, being $\Delta t (\equiv
\mathrm{t_{ind}})$ defined by the age where such steady-state is
reached.

\subsection{SSP luminosity evolving as a hat function plus a power law decay}

\label{sec:hat+pl}

\begin{figure}
\resizebox{\hsize}{!}{{\includegraphics{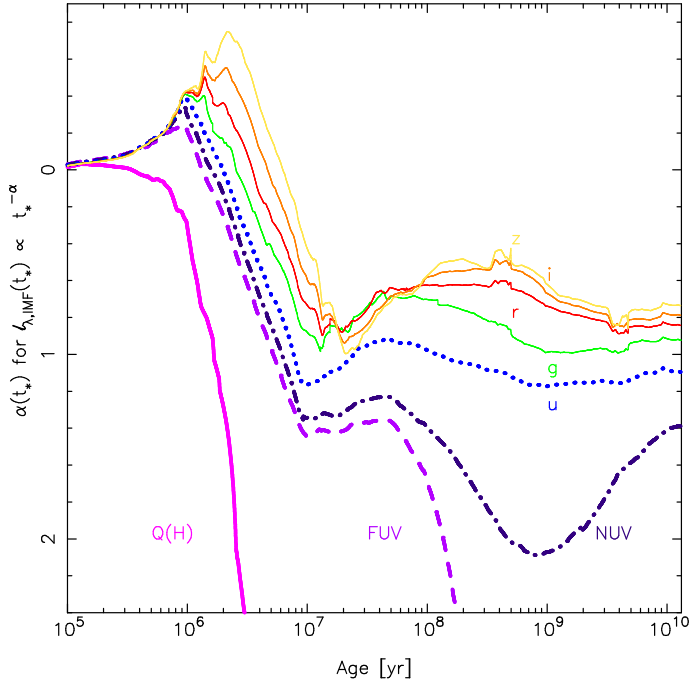}}}
\caption[]{Evolution of the slope of the approximation of SSP
  luminosity following power law evolution
  $\ell_{\lambda,\mathrm{IMF}}(t_*) \propto t_*^{-\alpha}$ (actually
  $\alpha(t_*)$) for different photometric bands obtained by the
  combination of different synthesis models (see Sect.~\ref{sec:eps}
  for details). The slope evolution of $Q(H)$ is only show up to 50
  Myr and {\sc{galex}}/FUV up to 200 Myr; in addition the slopes have
  been smoothed to represent the general aspect of the evolution. Note
  the non-standard orientation of the $y$-axis since refers to
  $\alpha$ values, whereas the slope is $-\alpha$.}
\label{fig:slopes}
\end{figure}

\begin{table*}
\begin{center}
\begin{tabular}{ l   |  r   |  rr  | rrrrr }
\hline index & $\alpha$ & $\left< t_* \right>_\lambda$ & \% at $\left<
t_* \right>_\lambda$ & $t_{\lambda,\mathrm{99\%}}$ &
$t_{\lambda,\mathrm{95\%}}$ & $t_{\lambda,\mathrm{90\%}}$ &
$t_{\lambda,\mathrm{80\%}}$ & $t_{\lambda,\mathrm{50\%}}$ \\ & &
$10^6$ yr & & $10^6$ yr & $10^6$ yr & $10^6$ yr & $10^6$ yr & $10^6$
yr \\ \hline generic $Q(H)$ & $>2.00$ & $<$ 13 & $>$89\% & $<$148 &
$<$ 30 & $<$ 15 & $<$ 7.5 & $<$ 3 \\ generic UV & 1.50 & 131 & 91\% &
3325 & 375 & 112 & 31 & 5 \\ generic U& 1.10 & 937 & 80\% & 11156 &
6187 & 3100 & 885 & 43\\ generic IR/V& 0.80 & 2549 & 67\% & 12457 &
10462 & 8337 & 5120 & 817\\ \hline \multicolumn{9}{c}{ $\alpha$ values
  from comparison with synthesis models in
  Sect.~\ref{sec:eps}}\\ \hline $Q(H)$ & 4.00 & 2 & 56\% & 9 & 5 & 4 &
3 & 2 \\ {\sc galex}/FUV & 1.55 & 101 & 91\% & 2386 & 254 & 80 & 24 &
5 \\ {\sc galex}/NUV & 1.50 & 131 & 91\% & 3325 & 375 & 112 & 31 & 5
\\ {\sc sdss}/$u$ & 1.07 & 1060 & 79\% & 11401 & 6824 & 3683 & 1156 &
58 \\ {\sc sdss}/$g$ & 0.88 & 2055 & 71\% & 12284 & 9753 & 7243 & 3861
& 413 \\ {\sc sdss}/$r$ & 0.75 & 2865 & 65\% & 12534 & 10796 & 8885 &
5834 & 1157 \\ {\sc sdss}/$i$ & 0.72 & 3055 & 64\% & 12573 & 10965 &
9170 & 6228 & 1388 \\ {\sc sdss}/$z$ & 0.66 & 3430 & 62\% & 12636 &
11246 & 9656 & 6936 & 1890 \\ \hline
\end{tabular}
\end{center}
\caption[]{Values of the slope of the SSP luminosity evolution
  $\ell_{\lambda,\mathrm{IMF}}(t_*)$ when modeled as a power law
  $\alpha$ (see below), the mean age of
  $\ell_{\lambda,\mathrm{IMF}}(t_*)$ denoted as $\left< t_*
  \right>_\lambda$, the percentage of the sensitivity of
  $\ell_{\lambda,\mathrm{IMF}}(t_*)$ in the 0 to $\left< t_*
  \right>_\lambda$ age range, and the ages where the sensitivity to
  $\ell_{\lambda,\mathrm{IMF}}(t_*)$ reach a $x\%$ value of the total
  sensitivity, $t_{_\lambda,\mathrm{x\%}}$, for 99, 95, 90, 80 and
  50\% for the set of bands used in this work. The results assume that
  $\ell_{\lambda,\mathrm{IMF}}(t_*)$ is flat up to 3 Myr and follows a
  decreasing power law with exponent $\alpha$ for larger ages up to
  $\mathrm{t_{age}} = 13 ~\mathrm{Gyr}$.  The upper part of the table
  shows the generic $\alpha$ values used in this section for different
  bands guided by the results in Fig.~\ref{fig:slopes}.  The lower
  part of the table shows $\alpha$ values chosen a posteriori to
  roughly fit the results of detailed computations presented in
  Sect.~\ref{sec:eps} (table \ref{tab:tind}).  }
\label{tab:tindPL}
\end{table*}

{\bf 1.} Going forward, let us use a second still simplified but more
realistic functional form of the SSP luminosity evolution.  Assuming a
properly defined zero age main sequence, all stars increasing its
luminosity (at least in UV to IR wavelengths) up to the end of the
main sequence; hence any $\ell_{\lambda,\mathrm{IMF}}(t_*)$ will have
a first period with a slow increase of its luminosity at least up to
the age $\mathrm{t_{*,MS}}$ where more massive stars leave the main
sequence, which is typically 3 Myr. After that age, the presence of
post-main sequence evolutionary phases result in a more complicate
evolution. However, simple energetic arguments show that, in a quite
reasonable approximation, $\ell_{\lambda,\mathrm{IMF}}(t_*)$ evolves
as a declining power law. It is a classical result \citep{TG76,Buzz95}
still confirmed by comparisons with current synthesis models and
proven as an useful approximation \citep{Buzz05}. Just for simplicity,
let us assume that $\ell_{\lambda,\mathrm{IMF}}(t_*)$ is constant in
the interval $[0,\mathrm{t_{*,MS}}]$ and it evolves as
$t_*^{-\alpha}$, from $t_{*,\mathrm{MS}}$ up to any posible
$\mathrm{t_{age}}$, being $\ell_{\lambda,\mathrm{MS}}$ the luminosity
at $t_\mathrm{*,MS}$, so the evolution of such SSP luminosity is:

\begin{eqnarray}
\ell_{\lambda,\mathrm{IMF}}(t_*) &=
\begin{cases}
  \ell_{\lambda,\mathrm{MS}} \,\,  & \mathrm{for} \,\,\,Êt_* \leq \mathrm{t_{*,MS}}, \\
 \ell_{\lambda,\mathrm{MS}} \left(\frac{t_*}{\mathrm{t_{*,MS}}} \right)^{-\alpha} \,\, & \mathrm{for}Ê\,\,\,t_* >\mathrm{t_{*,MS}}.
 \end{cases}
\end{eqnarray}

As reference values, $\alpha$ is around or lower than 1 for
wavelengths larger than 3000\AA ~\citep{Buzz02}. Table 1 in
\cite{Buzz05} provides a detailed analysis including metallicity
effects showing that the slope flattens when metallicity decreases.
Also as reference, we show the evolution of $\alpha$ for different
photometric bands obtained by the combination of different synthesis
models (see Sect.~\ref{sec:eps} for details) in Fig.~\ref{fig:slopes}.
In practical terms we will consider in this section generic values of
$\alpha = 0.6$ to 0.9 as representation of IR/visible bands, and
$\alpha = 1.1, 1.5$ and larger than $2$ as generic representation of
$U$ band, UV bands, and the number of Hydrogen ionizing photons,
($Q(H)$, which is proportional to the emission luminosity of Hydrogen
recombination lines, as the H$\alpha$ emission line),
respectively. The numerical results obtained here assumes
$\mathrm{t_{*,MS}} = 3~\mathrm{Myr}$, and, when required,
$\mathrm{t_{age}} = 13~\mathrm{Gyr}$.  We show in
Tab.~\ref{tab:tindPL} a more detailed version of specific values of
$\alpha$ for different generic bands, and related quantities computed
using the present approximation and discussed in this section.  The
lower part of the table shows $\alpha$ values chosen a posteriori to
roughly fit the results when realistic synthesis models are used
(Sec.~\ref{sec:eps}, table \ref{tab:tind}).  Note that for $Q(H)$ we
use here a generic value of $\alpha > 2$ as a limit although a value
of $\alpha = 4$ would be more realistic nominal value.

{\bf 2.} The integral over time of such
$\ell_{\lambda,\mathrm{IMF}}(t_*)$ for $t > \mathrm{t_{*,MS}}$ can be
obtained analytically:

\begin{eqnarray}
\int_0^t \ell_{\lambda,\mathrm{IMF}}(t_*) \mathrm{d}t_* & =
\begin{cases}
 \frac{ \ell_{\lambda,\mathrm{MS}} \, \mathrm{t_{MS}}}{\alpha-1} \left( \alpha - \left( \frac{t}{\mathrm{t_{MS}}}\right)^{1 - \alpha }  \right)  & \text{for } \alpha \ne 1, \\
  \ell_{\lambda,\mathrm{MS}} \, \mathrm{t_{MS}}\left( 1 +  \ln \frac{{t}}{\mathrm{t_{MS}}}\right) & \text{for } \alpha = 1.
\end{cases}
\label{eq:asym}
\end{eqnarray}

Such integral only has an asymptote if $\alpha > 1$ with a value:

\begin{equation}
 \ell_\lambda^{\mathrm{asymp}} = \frac{\ell_{\lambda,\mathrm{MS}}\,
   \mathrm{t_{MS}} \,\alpha}{\alpha -1}.
\label{eq:asymlim}
\end{equation}

So, the luminosity at wavelengths/bands larger than 3000\AA ~never
reach an asymptotical value and the sensitivity of the SSP evolution o
the old SFH increases as the system evolves. This situation, when
translated to the statement that the the time integral of the SSP
luminosity reach an asymptotic value to define a reliable SFR index
situates $U$ in a limiting situation due to its metallicity dependence
\cite[][]{Buzz05}. Actually $U$ is considered as a reliable index by
some authors \cite[e.j.][ but see below]{Wiletal12,BBP14} but not by
others.

{\bf 3.} A direct comparison of Eqs.~\ref{eq:asym} and
\ref{eq:asymlim}, allows to evaluate the difference between the real
asymptotic value and the value obtained for any chosen $t$ provided
that $\alpha > 1$; hence to estimate possible values of
$\mathrm{t_{test}}$ where an asymptotical values have been actually
reached.  Evaluating Eq.~\ref{eq:asym} at $t_* = 13 \mathrm{Gyr}$ (1
Gyr, 100 Myr), and comparing with the asymptotic value we found that
the asymptotic values is underestimated by 39\% (51\%, 64\%) for
$\alpha =1.1$ corresponding to a generic $U$ band; in the case of
$\alpha = 1.5$ corresponging to FUV bands, the underestimate is 1\%
(4\%, 12\%). Finally the underestimation is less than 1 .5\% in the
three ages for $\alpha \geq 2$. So, with exception of $Q(\mathrm{H})$
based indices (and neglecting their flattering at older ages, see
sect.~\ref{sec:eps} below), asymptotical values are never reached
given the age of the Universe!  It is,

\begin{quote} 
to reach a steady state/quasi-asymptotic value, although desirable,
cannot be a strong requirement to define and calibrate SFR indices
since such asymptotic value is not reached even at cosmological time
scales; actually the more close we would be to the asymptotic value is
to use the one defined by the age of the system $\mathrm{t_{age}}$.
\end{quote}

Actually, a graphical inspection of the
$C_\mathrm{ind}(\mathrm{t_{test}})$ values quoted in appendix of
\citealt{BBP14} shows that, excluding $Q(H)$ and apparently
$C_\mathrm{FUV}$ at some metallicities, an asymptotical value of
$C_\mathrm{ind}(\mathrm{t_{test}})$ at $\mathrm{t_{test}} = 1
\mathrm{Gyr}$ has been not reached.

{\bf 4.} The fact that asymptotic values can not be reached implies
that we cannot define a characteristic time scale $\Delta t$ which
allows a direct transformation of $\left<{\cal{SFR}}\right>_{\lambda}$
in $\left< {\cal{SFR}}\right>_{\Delta t}$.  We stress that it is
implicit in the filter Nature gives us to infer the SFR (it is the
power law nature of the evolution of SSP luminosities).  However we
can try to obtain some usable summaries of
$\ell_{\lambda,\mathrm{IMF}}(t_*)$, which allows to obtain information
without take into account the functional form of
$\ell_{\lambda,\mathrm{IMF}}(t_*)$ explicitly; a similar problem
related with the characterization of photometric systems, or
probability distributions. A typical characterization is obtained by
the computing of cumulative distributions of the amount of flux
comprised from 0 up to a given $t_*$ value (examples are the way SFR
is calibrated; see also \citealt{Leretal12} Ê~or
\citealt{Johetal13}). In the following we show two alternative
approaches used in the literature.

{\bf 4.1.} The first one is to define a {\it mean} luminosity-weighted
age \cite[][]{Buzz02b,BBP14}, which can we defined taking into account
an assumed SFH;

\begin{equation}
\left< t_* \right>_{\lambda,\psi(t)} = \frac{\int_0^\mathrm{t_{age}}
  \, t_* \,\ell_{\lambda,\mathrm{IMF}}(t_*)\,
  \psi(\mathrm{t_{age}}-t_*)\,\mathrm{d}t_*}{\int_0^\mathrm{t_{age}}
  \,\ell_{\lambda,\mathrm{IMF}}(t_*) \, \psi(\mathrm{t_{age}}-t_*) \,
  \mathrm{d}t_*}.
\label{eq:teffsfh}
\end{equation}

It can be used as a measure of the mean age of the stars which
contributes to ${\cal{L}}_\lambda$ at different wavelengths, SFH and
IMF slopes \cite[e.g.][]{Buzz02b}.

Alternatively, it can be defined a characteristic weighted age of
$\ell_{\lambda,\mathrm{IMF}}(t_*)$ without considerations about the
SFH (or, equivalently at mathematical level, by assuming a constant
SFH over all the galaxy life-time),

\begin{equation}
\left< t_* \right>_{\lambda} = \frac{\int_0^\mathrm{t_{age}} t_*
  \,\ell_{\lambda,\mathrm{IMF}}(t_*)
  \mathrm{d}t_*}{\int_0^\mathrm{t_{age}}
  \,\ell_{\lambda,\mathrm{IMF}}(t_*) \mathrm{d}t_*},
\label{eq:teff}
\end{equation}

\noindent also used by \cite{Buzz02b}, and \cite{Leretal12} to study
the sensitivity of SFR to recent SFH variations, or by \cite{BBP14} to
investigate, by comparison with $\left< t_* \right>_{\lambda,\psi(t)}$
the stability of $C_\mathrm{ind}(\mathrm{t_{test}})$ as a function of
$\mathrm{t_{test}}$ and different SFHs. Using our power law
approximation, $\left< t_* \right>_{\lambda}$ can be obtained
analytically using Eq.\ref{eq:asym} easily, having values of 2.5 Gyr
(937, 131, $13$ Myr) for $\alpha = 0.8$ (1.1, 1.5, 2). Such values are
roughly in agreement with our expectations about $\Delta t$ based in
the stellar life-times which mainly contributes to different
wavelengths.

The use of a simplified $\ell_{\lambda,\mathrm{IMF}}(t_*)$ also allows
to compute easily the amount of sensitivity up to
$\left<t_{*}\right>_{\lambda}$, being the results shown in
Table~\ref{tab:tindPL}.  Using first principles, given the L-shape
nature of $\ell_{\lambda,\mathrm{IMF}}(t_*)$, we can assure that at
least 50\% of the sensitivity to the SFH is concentrated at ages equal
or lower than $\left< t_* \right>_{\lambda}$ for any band (including
optical ones), although the value depends on $\alpha$ being a maximum
value reached at $\alpha \sim 1.67$.  So, as a remark, $\left< t_*
\right>_\lambda$ it provides valuable information, but it does not
provide neither a cut-off in $\ell_{\lambda,\mathrm{IMF}}(t_*)$, nor a
characteristic time over the recent SFH is averaged.

{\bf 4.2.} A second characterization of the evolution of SSP
luminosities is to compute the ages $t_{\lambda,x\%}$ where the
sensitivity of $\ell_{\lambda,\mathrm{IMF}}(t_*)$ to any SFH comprises
$x\%$ of the total sensitivity, which is obtained solving:

\begin{equation}
\int_0^{t_{\lambda,x\%}} \,\ell_{\lambda,\mathrm{IMF}}(t_*)
\mathrm{d}t_* = \frac{x}{100} \, \int_0^{\mathrm{t_{age}}}
\,\ell_{\lambda,\mathrm{IMF}}(t_*) \mathrm{d}t_*,
\label{eq:txp.c.}
\end{equation}

An advantage of $t_{\lambda,x\%}$ is that provide a more quantitative
information that $\left< t_* \right>_{\lambda}$. Again, it cannot be
taken as a face-on value of $\Delta t$ but, at least, provide
information about how much of the sensitivity of the curve would be
affected by the old component of the SFH.

Using our simplified evolution of SSP luminosities, we obtain values
of $t_{\lambda,\mathrm{80\%}}$ ($t_{\lambda,\mathrm{95\%}}$) of 5.1
Gyr (10.5 Gyr) for $\alpha =0.8$ which are {\sc sdss}/$g$ filters; 885
Myr (6.2 Gyr) for $\alpha =1.1$ or $U$ band; 31 Myr (375 Myr) for
$\alpha = 1.5$ or UV filters, and 7.5 Myr (30 Myr) for $\alpha =2$,
i.e. the ionizing flux (c.f, Tab~\ref{tab:tindPL}).  Values obtained
using detailed synthesis models results are shown in
Tab.~\ref{tab:tind} and discused in Sec.~\ref{sec:eps}.  Note that,
given that $t_{\lambda,100\%} = \mathrm{t_{age}}$ by construction,
each $t_{\lambda,x\%}$ is also a measure about how far/close we are to
the physical limiting value when $t_{\lambda,x\%}$ is used to define
$C_\mathrm{ind}$.  Of course, as in the case of $\left< t_*
\right>_{\lambda,\psi(t)}$, the definition can be extended to any SFH
\cite[see][ as an example]{Johetal13}.

{\bf 5.} As a summary of results, we have seen how our expectations
about SFR inferences had been downgraded: we have first relaxed out
expectations of obtain $\psi(\mathrm{t_{now}})$ to obtain an averaged
over a defined $\Delta t$, $\left< {\cal{SFR}}\right>_{\Delta t}$. But
given the nature of the integrated luminosity, we have downgrade again
to obtain a $\left<{\cal{SFR}}\right>_{\lambda}$ where a single time
scale over the SFR has been averaged can no properly defined. The most
we can obtain is the sensitivity of the given luminosity to the recent
and old components of the global SFH. As a collateral result is that
such kind of information can be obtained for any luminosity (not only
the standard ones used a SFH indices). When applied to optical fluxes,
we obtain that 50\% of the sensitivity of the integrated luminosity is
concentrated at ages lower than 2 Gyr, so such wavelengths still
contains a valuable information about the recent (lower than 2 Gyr)
SFH of the system. Such wavelengths can be used to constraint the
quality of SFR inferences obtained by bona fide indices as we will se
below.

\subsection{SSP luminosity evolution
computed by synthesis models}
\label{sec:eps}

\begin{figure}
\resizebox{\hsize}{!}{\includegraphics{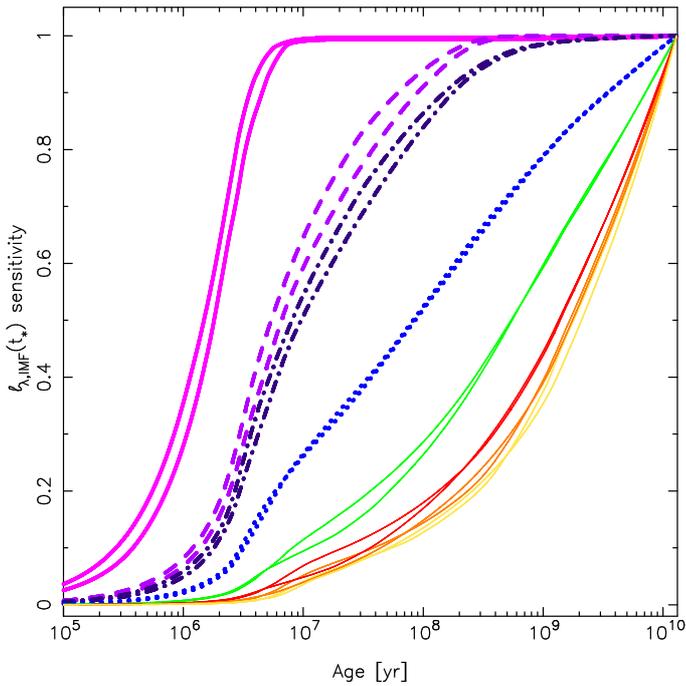}}
\caption[]{Evolution of the sensitivity of the SSP luminosity
  $\ell_{\lambda,\mathrm{IMF}}(t_*)$ with the age using the upper and
  lower envelopes of SSP results (see text); the age (actually age
  range) corresponding to a given sensibility, $t_{\lambda,x\%}$ can
  be directly compared with the limits quoted in Table~\ref{tab:tind}
  for the different luminosities. In ascending ages each set of two
  curves correspond to $Q(H)$, {\sc galex}/FUV and NUV, and {\sc
    sdss}/$u$, $g$, $r$, $i$ and $z$.  These curves can be also
  interpreted as the evolution of synthesis models under a constant
  SFR assumption, modulus the normalization factor.}
\label{fig:CSFR}
\end{figure}

\begin{table*}
\begin{tabular}{l l l l l l l}
\hline index & $t_{\lambda,99\%}$ & $t_{\lambda,95\%}$ &
$t_{\lambda,90\%}$ & $t_{\lambda,80\%}$ & $t_{\lambda,50\%}$ & $-\log
C_\lambda$\\ & $10^6$ yr & $10^6$ yr & $10^6$ yr & $10^6$ yr & $10^6$
yr & \\ \hline $Q(H)$ & 8.7 ( 7.6- 9.5)& 5.3 ( 4.5- 5.7)& 4.2 ( 3.6-
4.6) & 3.1 ( 2.8- 3.3)& 1.7 ( 1.5- 1.8)& 52.93 ( 52.82- 53.02) \\ FUV
& 303 ( 266- 330)& 141 ( 117- 156)& 77 ( 61- 89)& 31 ( 24- 37)& 6 ( 5-
6)& 39.99 ( 39.93- 40.05) \\ NUV & 1481 ( 1445- 1508)& 335 ( 310-
353)& 166 ( 149- 177)& 64 ( 54- 72)& 8 ( 8- 9)& 39.63 ( 39.55- 39.69)
\\ \hline index & $t_{\lambda,99\%}$ & $t_{\lambda,95\%}$ &
$t_{\lambda,90\%}$ & $t_{\lambda,80\%}$ & $t_{\lambda,50\%}$ & $-\log
C_\lambda$\\ & $10^9$ yr & $10^9$ yr & $10^9$ yr & $10^9$ yr & $10^9$
yr & \\ \hline L$_\mathrm{bol}$ & 12.31 ( 12.27- 12.32)& 9.67 ( 9.65-
9.71)& 6.96 ( 6.96- 6.96)& 3.30 ( 3.31- 3.29)& 0.13 ( 0.10- 0.15)&
43.68 ( 43.62- 43.74) \\ $u$ & 11.29 ( 11.31- 11.29)& 6.60 ( 6.60-
6.61)& 3.47 ( 3.45- 3.49)& 1.12 ( 1.13- 1.12)& 0.08 ( 0.09- 0.08)&
39.36 ( 39.29- 39.42) \\ $g$ & 12.25 ( 12.29- 12.24)& 9.65 ( 9.70-
9.59)& 7.15 ( 7.17- 7.13)& 3.81 ( 3.78- 3.84)& 0.55 ( 0.55- 0.54)&
39.48 ( 39.42- 39.54) \\ $r$ & 12.52 ( 12.53- 12.50)& 10.69 ( 10.77-
10.63)& 8.75 ( 8.85- 8.68)& 5.79 ( 5.84- 5.75)& 1.37 ( 1.35- 1.39)&
39.40 ( 39.32- 39.46) \\ $i$ & 12.59 ( 12.60- 12.57)& 10.95 ( 11.03-
10.89)& 9.16 ( 9.29- 9.07)& 6.31 ( 6.44- 6.23)& 1.70 ( 1.76- 1.67)&
39.31 ( 39.23- 39.38) \\ $z$ & 12.62 ( 12.64- 12.60)& 11.12 ( 11.20-
11.06)& 9.43 ( 9.60- 9.31)& 6.67 ( 6.90- 6.51)& 1.91 ( 2.09- 1.79)&
39.25 ( 39.15- 39.33) \\ \hline
\end{tabular}
\caption[]{ Ages where the sensitivity to
  $\ell_{\lambda,\mathrm{IMF}}(t_*)$ reach a $x\%$ value of the total
  sensitivity, $t_{\lambda,x\%}$, for 99, 95, 90, 80 and 50\% and all
  luminosities used in this work. Last column is the $C_\lambda$ value
  as defined in Eq.~\ref{eq:meanSFRok}. The units of $C_\lambda$ are
  in erg s$^{-1}$ \AA$^{-1}$ M$_\odot^{-1}$ yr for the standard
  photometric systems and photons s$^{-1}$ M$_\odot^{-1}$ yr for
  $Q(H)$. Values in parenthesis corresponds to the use of upper and
  lower envelope of $\ell_{\lambda,\mathrm{IMF}}(t_*)$ obtained from
  our calibration of SSP models.  }
\label{tab:tind}
\end{table*}

{\bf 1.} Once we have used suitable examples to manage the
characterization of the evolution of SSP luminosities and estimate
some numbers based in an approximate formulation of the problem, let's
examine what the explicit computation of
$\ell_{\lambda,\mathrm{IMF}}(t_*)$ provides. Inevitably, it implies
the use of evolutionary synthesis codes to perform the detailed
numerical computations, and the result becomes dependent of the
details of the used code (interpolations, numerical methods,
ingredients). To overcome such situation, we have compiled the results
of 13 different synthesis codes/stellar population
results\footnote{The used modelas are: {\sc starbust99}
  \cite[][]{sb99,Leietal14}, {\sc galev} \cite[]{galev}, {\sc galaxev}
  \cite[][version 2012]{bc03}, {\sc pegase2.0}
  \cite[][]{pegase,pegase2}, {\sc popstar}
  \cite[][]{popstar,popstar2,popstar3}, {\sc fsps}
  \cite[][]{fsps,fsps2,fsps3}, {\sc galadriel} \cite[][]{galadriel},
       {\sc bpass} \cite[][ in its single star
         version]{bpass1,bpass2}, {\sc sed@}
       \cite[][]{MHK91,sed1,sed2}, models provided by C. Maraston
       \cite[][]{claudia1,claudia2} and A. Buzzoni \cite[][]{Buzz89}
       with different Horizontal brach morphologies, models from {\sc
         batsi} web server including different $\alpha$-enhancement
       factors \cite[][]{batsi,batsi2,batsi3}, and models from {\sc
         CMD 2.0} web server \cite[Padova
         models,][]{Giretal02,Giretal08,Maretal08}.  The web address
       of the models can be found in {\tt http://sedfitting.org}.}
which results are public available. The models includes different
atmosphere models\footnote{Atmosphere models includes: grids by
  \cite{K91,CGK97}, different versions of {\sc basel} libraries
  \citep{basel1,basel2,basel3} for normal stars, the grids by
  \cite{SNC02}, \cite{SLG92}, and {\sc CoStar} \cite[][]{costar} for
  massive and WR stars, and Planck functions and \cite{rauch} models
  for white dwarfs (WD).} and evolutionary tracks/isochrone
sets\footnote{The tracks/isochrones used by the different models
  models are: Geneva tracks \cite[][]{gen1}, Padova tracks
  \cite[][]{pad1,pad2,pad3}, {\sc batsi} tracks
  \cite[][]{bat1,bat2,bat3,batsi,batsi2,batsi3}, and
  \cite{wd1,wd2,wd3,wd4} for post-AGB/WD evolution.}; neither
binaries, rotation or evolution with enhanced mass loss rates has been
considered.  All models assume metallicities between 0.020 and 0.019,
and use (or had been transformed to) a \cite{Sal55} IMF in the mass
range 0.01-100 M$_\odot$ (the impact of variations of the IMF slope at
low mass does not affects the present results; we note that some
models has been computed with a $m_\mathrm{up} = 120
\mathrm{M}_\odot$, which has been taken into account in the censorship
process, see below). No nebular continuum neither emission lines or
attenuation effects have been taken into account.

We have use the computed low resolution spectral energy distribution
(SED) provided by each model, to obtain the fluxes in $Q(H)$, {\sc
  galex}/FUV and NUV bands, and {\sc sdss}/$u$, $g$, $r$, $i$, and $z$
bands\footnote{Filter transmission curves has been taken from the
  spanish virtual observatory, SVO, server at {\tt
    http://svo2.cab.inta-csic.es/theory/fps3/}}; we have crosschecked
that our results are coincident with the fluxes in these bands when
provided by the modeler (and exception are results from {\sc cmd 2.0}
server which provides the fluxes in all considered bands, except
$Q(H)$, but not the corresponding SEDs).  After a censorship
process\footnote{Roughly, we discard the age ranges of models which
  shows serious discrepant results from the overall behavior of the
  ensemble, specially when such discrepancy is reported by the absence
  of particular evolutionary phases, or when the discrepant age range
  is outside the modeler expertise (which inferred from the age range
  where modelers shows their results in refereed journals).}, we have
obtain the upper and lower envelopes from the censored set of
models. Then, we define a reference SSP luminosity evolutuion
$\ell_{\lambda,\mathrm{IMF}}(t_*)$ by the linear mean value between
both envelopes.  Details are presented in a companion paper (Cervi\~no
et al. 2016 in prep.).

{\bf 2.} The resulting ages where the sensitivity to the luminosity
evolution of the SSP $\ell_{\lambda,\mathrm{IMF}}(t_*)$ reach a $x\%$
value of the total sensitivity, $t_{\lambda,x\%}$, and $-\log
C_\lambda$ values obtained for $\mathrm{t_{age}} = 13 ~\mathrm{Gyr}$
are shown in Table \ref{tab:tind}.  Nominal values corresponds to the
reference model and the values corresponding to the upper and lower
envelopes (i.e. the admisible range for where any public model is
enclosed) is quoted in brackets.  The age limits quoted in Table
\ref{tab:tind} can be also obtained from Fig.~\ref{fig:CSFR} where we
show the evolution of the sensitivity of the SSP luminosity
$\ell_{\lambda,\mathrm{IMF}}(t_*)$ with the age using the upper and
lower envelopes of SSP results (being each of the envelopes normalized
to its corresponding value).  This curves can be also interpreted as
the evolution of synthesis models under a constant SFR assumption,
modulus the normalization factor. The figure shows how the dispersion
in the results of different synthesis models and models ingredients
propagates in $t_{\lambda,x\%}$ values (or in the resulting evolution
under a constant SFR assumption).

The values obtained in Table \ref{tab:tind} are comparable with
$t_{\lambda,90\%}$ provided in table 1 of \cite{KE12} based in
computations by \cite{Haoetal11} and \cite{Muretal11}, although we
obtain lower $t_{\lambda,90\%}$ values. This is a surprising result
given that we use a quite larger $\mathrm{t_{test}}$; although
$\mathrm{t_{age}}$ and our SSP calibration includes the emission of
stellar components. which are not included in the models used by
\cite{Haoetal11,Muretal11,KE12}. This difference would be due to the
use of \cite{Meyetal94} evolutionary tracks with enhanced mass loss
rates by the mentioned authors, the default in {\sc starbust99}
previous the release including rotation, which are not included in our
censored calibration (see Cervi\~no et al. 2016 in prep. for more
details).

The variability due to the use of different synthesis models in our
compilation quoted in Table \ref{tab:tind} is quite lower than the
20\% usually quote in the literature. However, such scatter
corresponds to an optimistic situation since our compilation is
restricted to the evolutionary tracks used in common synthesis
codes. A detailed analysis of possible uncertainties due to
evolutionary tracks which are not included in our compilation can be
found in \cite{MP13}. In addition, the compilation only include solar
metallicity models, so, again, the quoted uncertainties are lower
limits since does not consider metallicity variations.

\begin{figure*}
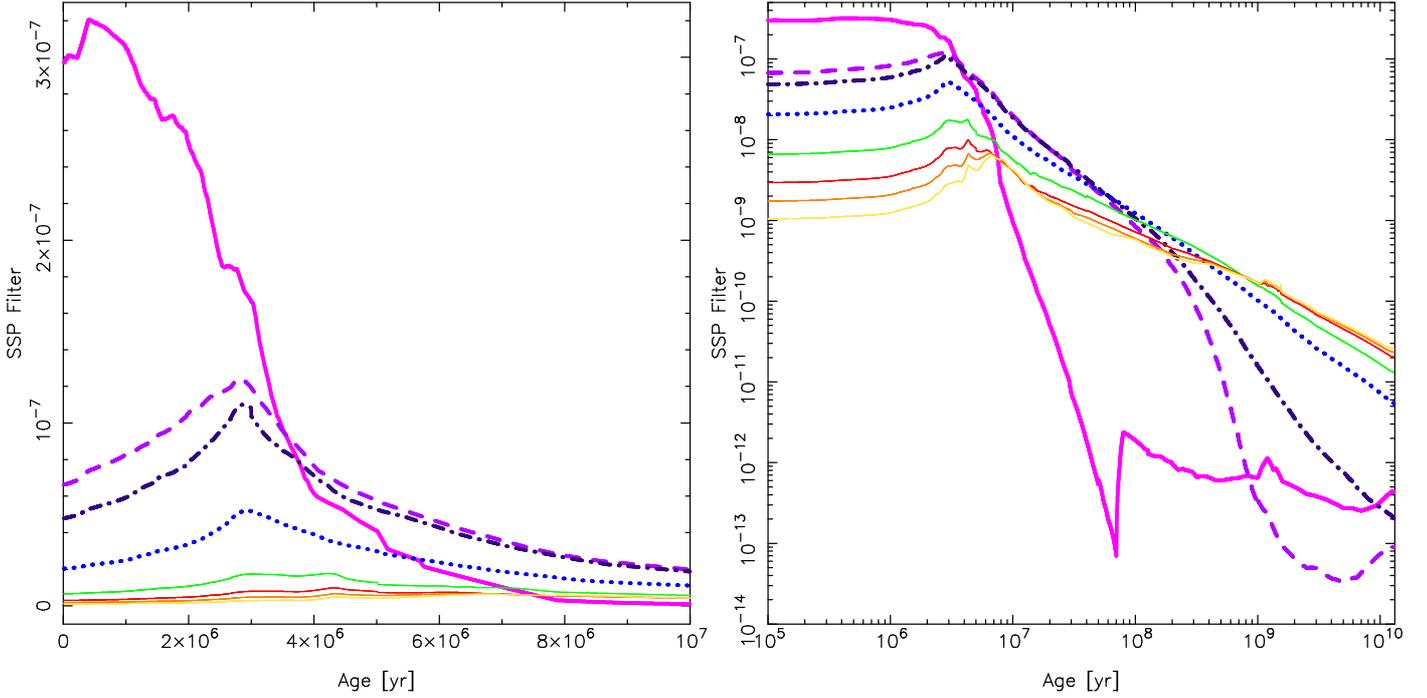

\resizebox{\hsize}{!}{{\includegraphics{CBH_fig3a.eps}}{\includegraphics{CBH_fig3b.eps}}}
\caption[]{ SSP luminosity evolution
  $\ell_{\lambda,\mathrm{IMF}}(t_*)$ as SFH sensitivity curve
  (i.e. once normalized to the integral of the SSP over the age of the
  system, 13 Gyr in our case).  The left panel shows the sensitivity
  curve in linear scale from 0 to 10$^7$ yr, and the right panel the
  sensitivity curve in log-log scale in the whole age range. In
  descending order at young ages the curves correspond to $Q(H)$, {\sc
    galex}/FUV and NUV, and {\sc sdss}/$u$, $g$, $r$, $i$ and $z$.}
\label{fig:sfltref}
\end{figure*}

{\bf 3.} Figure \ref{fig:sfltref} shows the
$\ell_{\lambda,\mathrm{IMF}}(t_*)$ sensitivity curves once normalized
to its integral over 13 Gyr, which is the transmission over which the
SFH is seen by the corresponding luminosity. The figure allows to
compare directly with the sensitivity to the SFH for each possible
integrated luminosity independently if it is used as a recent SFH
proxy or not. To simplify the discussion, we have only used the
reference model described before.  The left panel in the figure shows
the sensitivity in linear scale form 0 to 10 Myr, and right panel
shows the sensitivity in logarithm scale in the whole age range. In
the following paragraphs we compare the four groups of indices with
different behavior, which are $Q(H)$, UV indices, $U$ ({\sc
  sdss}/$u$), and optical/IR indices.

$Q(H)$ is clearly the most sensitive index to the younger component of
the SFH. Even more, the sensitivity peaks at ages lower than 1 Myr,
hence, in first approximation, it almost reproduce the present value
of the SFH. In addition, its sensitivity to the recent SFH
($\mathrm{t_{now}} - 3 \mathrm{Myr}$) is about a factor 3 larger than
any other index. It is the less sensitive index to the SFH at ages
$\mathrm{t_{now}} - 10 \mathrm{Myr}$ up to ages older than 1 Gyr,
where the sensitivity of {\sc galex}/FUV is lower.  The dynamic range
of the sensitivity to the SFH at different ages covers 6 decades (more
than 3 decades in the first 10 Myr), hence, it is quite
stable\footnote{Numerical computations shows that, assuming
  $\mathrm{t_{age}} = 13 \mathrm{Gyr}$, the old component of
  exponential decay and delayed SFH with $\tau > 3 \mathrm{Gyr}$
  affects the index in less than 10\% (Cervi\~no et al. 2016b in
  preparation).} to large scale variations in the SFH at ages older
than 50 Myr. In a relative comparison with the other indices
(i.e. where the different sensitivities crosses each other), $Q(H)$ is
more sensitive to the SFH at ages lower $\sim$4 Myr than {\sc galex}
filters, $\sim$5 Myr than $u$ and $\sim$7 Myr than optical bands.

The indices based in the UV, {\sc galex}/FUV and NUV, have a quite
similar transmission, although {\sc galex}/FUV is a bit more sensitive
to the young component up to ages around 8 Myr than {\sc galex}/NUV,
and {\sc galex}/NUV is more sensitive than {\sc galex}/FUV for the SFH
at ages longer than 100 Myr. The peak of the sensitivity is around 3
Myr (the value of $\mathrm{t_{MS}}$ at the given metallicity), being
the sensitivity of both indices broader than $Q(H)$ and extending with
an apreciable sensitivity for ages longer than 10 Myr. Both indices
have almost equivalent sensitivity to the SFH in the range 8 to
$\sim$50 Myr. At older ages, and specially at ages older than $\sim$
300 Myr the sensitivity of {\sc galex}/FUV drops abruptly, whereas the
one of {\sc galex}/NUV declines more smoothly. The dynamic range of
the sensitivity at different ages covers almost 5 decades (more than 3
decades in the first 500 Myr), and, as in the case of $Q(H)$, both
indices are quite robust to large scale variations of the SFH,
although at a time scale much more larger that the one associated to
$Q(H)$.

The case of $U$ band is and intermediate case between optical and UV
bands. It is about a factor 2 less sensitive to the recent SFH than UV
filters but still a factor 2 lager than $g$; however, the sensitivity
to the SFH after 50 Myr is larger than the UV bands (reaching factors
larger than 10 at ages large than 2-3 Gyr). So, although it looks to
works correctly as a recent SFR index using the standard methodology
when tested over sort time-scales (i.e. $\mathrm{t_{test}}$ up to
$\sim$ 100 Myr), it behaves more like optical colors a larger ages.
Actually, the slope of the sensitivity with time is quite similar to
$-1$, which is the limiting case where the sensitivity to young and
old components of the SFH are similar. The dynamic range of the
sensitivity is a bit larger than 3 decades over the whole age range,
and, as quoted before, more sensitive to large scale variations of the
SFH than the previous indexes.

Larger wavelengths ($g$, $r$, $i$, and $z$ bands), still shows a
important sensitivity to the recent SFH, however, their dynamic range
is lower than 3 decades, hence, much more affected by large scale
variations on the SFH. In addition, the sensitivity curves of all
optical bands intercept each other near $1~\mathrm{Gyr}$. Among them,
the sensitivity of $r$, $i$ and $z$ bands are quite similar, which
implies in a first approximation that they would provide redundant
information in any SFH inference, specially after the first 10 Myr.

\subsubsection{Relative time-scales and $\left<{\cal{SFR}}\right>_{\lambda}$ corrections}

\begin{table*}
\begin{tabular}{r | r r | r r | r r | r r | r r | r r}
\hline & ${\cal B} -$ NUV & $t_{*,{\cal B}-{\mathrm{NUV}}}$ & ${\cal
  B} - u$ & $t_{*,{\cal B}-u}$ & ${\cal B} - g$ & $t_{*,{\cal B}-g}$ &
${\cal B} - r$ & $t_{*,{\cal B}-r}$ & ${\cal B} - i$ & $t_{*,{\cal
    B}-i}$ & ${\cal B} - z$ & $t_{*,{\cal B}-z}$ \\ & [AB] & [Myr] &
       [AB] & [Myr] & [AB] & [Myr] & [AB] & [Myr] & [AB] & [Myr] &
       [AB] & [Myr] \\ \hline FUV$ - {\cal R}$ & $-0.02 $ & (7 - 50) &
       0.25 & (40-60) & 1.16 & (80) & 1.55 & ($\sim$120) & 1.74 &
       ($\sim$150) & 1.98 & ($\sim$150)\\ NUV$ - {\cal R}$ & & & 0.27
       & (40-50) & 1.18 & (100) & 1.57 & ($\sim$180) & 1.76 &
       ($\sim$200) & 2.00 & ($\sim$200)\\ $u - {\cal R}$ & & & & &
       0.91 & ($\sim$180) & 1.30 & ($\sim$400) & 1.49 & ($\sim$400) &
       1.73 & ($\sim$400)\\ $g - {\cal R}$ & & & & & & & 0.39 &
       ($\sim$700) & 0.59 & ($\sim$700) & 0.82 & ($\sim$700)\\ $r -
       {\cal R}$ & & & & & & & & & 0.19 & (400 - 1000) & 0.43 & (400 -
       1000)\\ $i - {\cal R}$ & & & & & & & & & & & 0.24 & (400 -
       4000)\\ \hline
\end{tabular}
\caption[]{Colors obtained from the normalized
  $\left<{\cal{SFR}}\right>_{\lambda}$ calibration in magnitudes in
  the AB system. The age where each sensitivity curve cross each other
  in Myr units is quoted in parenthesis. In the case of $r$, $i$ and
  $z$ combination of colors there is an additional crossing age in the
  7- 13 Myr range not quoted in the table.}
\label{tab:AB}
\end{table*}

{\bf 1.} In the previous section we showed the difficulties to define
any characteristic time scale $\Delta t$ which allows transform an
observed $\left<{\cal{SFR}}\right>_{\lambda}$ into
$\left<{\cal{SFR}}\right>_{\Delta t}$ or, at least obtain an age
interval over the SFH has been averaged. We can choose a
characteristic time scales associated with the evolution of SSP
luminosities $\ell_{\lambda,\mathrm{IMF}}(t_*)$ (e.g. $\left< t_*
\right>_\lambda$, any $t_{\lambda,x\%}$ or any other related time
scale), but they do not provides directly the time range over the
actual SFH is averaged neither $\left< t_* \right>_{\lambda,\psi}$, or
$t_{\lambda,x\%,\psi}$, which depend on the unknown functional form of
the overall SFH.

However, by the comparison of the $\left<{\cal{SFR}}\right>_{\lambda}$
obtained for different indices (including optical ones), we can obtain
relative time scales of the SFH whatever its functional form. It is,
we cannot define the time interval over $\psi(t)$ is averaged, but we
can establish some characteristic times which, once compared with an
associated color, allows to establish the relative strength of
$\psi(t)$ after and before such time. As result, although we can not
correct $\left<{\cal{SFR}}\right>_{\lambda}$ to obtain
$\left<{\cal{SFR}}\right>_{\Delta t}$, we can establish if
$t_{\lambda,x\%,\psi}$ (which is unknown) is larger or lower than
$t_{\lambda,x\%}$.  In the following we assume the general result that
the sensitivity to the recent SFR increases at lower wavelengths.

{\bf 2.} Relative time scales are given by the intersection of the
different transmission curves: Let us assume two indices $C_{{\cal
    B}}$ and $C_{{\cal R}}$ where $\cal B$ and $\cal R$ refers to
bluest or redder bands used to define the color, or in terms of the
transmission curves, more sensitive to the young ($\cal B$) or old
($\cal R$) component of the SFH. First, let us define a reference
color $({\cal B} - {\cal R})_\mathrm{ref}$ obtained from the
corresponding $C_\lambda$ values (i.e. obtained at
$\mathrm{t_{age}}$). Second, let be $t_{*,{\cal B}{\cal R}}$ the
intersection age of the two sensitivity curves. Given that $\psi(t)$
is independent of the transmission curves, an extinction-corrected
observed color bluer than $({\cal B} - {\cal R})_\mathrm{ref}$ implies
that $\psi(t)$ have a larger contribution in the age region where the
blue index is more sensitive, it is, at ages lower than $t_{*,{\cal
    B}{\cal R}}$. In such situation, we can also assure that any of
the time scales $\left< t_* \right>_\lambda$ or $t_{\lambda,x\%}$ are
upper values of the actual $\left< t_* \right>_{\lambda,\psi(t)}$ or
$t_{\lambda,x\%,\psi(t)}$ values. It is, from the variation of the
color $({\cal B} - {\cal R})$ with respect to $({\cal B} - {\cal
  R})_\mathrm{ref}$ we can obtain information about the relation
between $\left< t_* \right>_\lambda$ (obtained theoretically) and
$\left< t_* \right>_{\lambda,\psi(t)}$ (the cuantity we are interested
in).

This case can be viewed as the comparison of the colors obtained from
a constant SFH over all the possible age range with any other possible
SFH. The improvement is that we have take advantage of the functional
form of the different normalized $\ell_{\lambda,\mathrm{IMF}}(t_*)$
curves and their intersection in the time axis to characterize the
deviations from a constant SFH.

Let us illustrate it with an example: $Q(H)$ is not directly an
observable but it is directly proportional to the H$\alpha$ emission
line with a conversion factor of $1.36\times10^{-12}$, (assuming Case
B recombination and no scape of ionizing photons, hence an upper limit
of $L(\mathrm{H}\alpha$)).  Using the flux in $r$ band as a
representation of the continuum near H$\alpha$, the resulting
equivalent width of H$\alpha$ in emission obtained from the respective
$C_{Q(H)}$ and $C_{r}$ values is EW(H$\alpha$) $\sim$ 45 \AA. Note
that in this computation the value of $r$ is a lower limit since we
are not considering nebular contribution to $r$ \cite[which is around
  40\% at young ages][]{MHK91}, so 45 \AA ~is a maximum value.  Since
the sensitivity curve of $Q(H)$ and $r$ intercepts at around 7 Myr,
the SFH in a system with EW(H$\alpha$) $> 45$ \AA, the actual SFH must
be stronger (in reference to a constant SFH) in the last 7 Myr. A
larger value of EW(H$\alpha$) implies that recent SFH is more
concentrated at younger ages, hence the mean luminosity weighted age
associated to the actual SFH) $\left< t_* \right>_{\psi(t)}$ is lower
than the mean luminosity weighted age associated to a constant SFH)
$\left< t_* \right>$, hence the recent SFH is bursty-like (at least in
first approximation).  However, the inverse reasoning of a recent SFH
extending in time for ages larger than 7 Myr if EW(H$\alpha$) $< 45$
\AA ~is not true since it can be due to the enhanced of the $r$ due to
nebular emission we have not consider, the leaking of ionizing
photons, and/or in combination that the SFH at ages larger than 7 Myr
is more relevant to the integrated $L(\mathrm{H}\alpha$).  Whatever
the case, the EW(H$\alpha$) value and the normalized
$\ell_{\lambda,\mathrm{IMF}}(t_*)$ curves provide additional
information about the recent SFH which helps to interpret the quantity
$\left<{\cal{SFR}}\right>_{Q(H)}$ independently of the SFH itself.
Equivalently FUV-NUV colors larger (or lower) than -0.02 or NUV-$r$
larger or lower than 1.57 provides additional constraints about the
time scales around 7-50 Myr and 140 Myr respectively
(c.f. Tab.~\ref{tab:AB}).

{\bf 3.} In the previous paragraph we had focused in provide a
time-scale to the $\left<{\cal{SFR}}\right>_{\lambda}$ obtained from
data of a single system. In the case of a large set of systems
(e.g. survey studies), the principal interest is not the time scale
associated to the $\left<{\cal{SFR}}\right>_{\lambda}$ in each system,
but to the comparison of $\left<{\cal{SFR}}\right>_{{\Delta t}}$ where
$\Delta t$ is equal (or at least similar) for all the systems in the
set. In such case, the comparison of the observed color $({\cal B} -
{\cal R})$ with respect to $({\cal B} - {\cal R})_\mathrm{ref}$
provide a hints about the correction to transform
$\left<{\cal{SFR}}\right>_{\lambda}$ in
$\left<{\cal{SFR}}\right>_{{\Delta t}}$.

{\bf 4.} However, although the idea is formally correct, this method
only provide first order time scales. As an example the {\sc
  galex}/FUV and NUV sensitivities crosses each other nominally at 17
Myr, but the sensitivity is almost identical (with variations lower
than $\pm 10\%$) in the age range from 7 to 50 Myr\footnote{These
  numbers has been obtained without take into consideration the
  uncertainties in the our calibration of synthesis models, which
  introduces an additional scatter in the reliable time scales}. In
addition, the present sensitivity curves have been obtained assuming
that all stars formed in over the last 13 Gyr has solar metallicity,
which neglects metallicity evolution of different
populations. Finally, we had not consider extinction effects which
affects the results of SFR inferences and which had been studied by
different authors.  Being quoted the previous cautions, we show in
Table \ref{tab:AB} the $({\cal B}-{\cal R})_\mathrm{ref}$ colors
associated to the different $C_\mathrm{ind}$ values of table
\ref{tab:tind} when expressed AB magnitudes, and the approximate ages
(obtained by eye-inspection of Fig.~\ref{fig:sfltref}) where the
sensitivity curves cross each other.

{\bf 5.} Finally, we stress that present results are independent of
the SFH, and apply also to the extreme SFH of instantaneous burst of
star formation. In the case of EW(H$\alpha$), an EW(H$\alpha$) $> 45$
\AA ~roughly corresponds to a burst (i.e. SSP) older than 7 Myr. A
direct implication is that, in practice, we can interpret any fit of
colors obtained from SFR indices to SSP results as a hint about the
different time-scales each index applies. Although outside the scope
of this paper, such alternative vision about what provide a SSP fit,
even in the case that we know a priori that our studied system is not
a single burst of star formation, can be potentially exploited in SFH
inferences obtained from the integrated spectra/photometry of any
stellar system.

\section{Discussion by comparison with other works}
\label{sec:disc}

The principal result of this work is a change of perspective about
what is obtained in recent SFH inferences. This result has not a great
impact on the final values of the standard SFR calibrations ($Q(H)$
and UV indices) which are only affected in a few percent, but it
clearly affects the case of the $U$ band and allows to introduce
optical colors as a cross-checking about the time scales associated to
SFR inferences. Although we have obtained some numbers, our approach
is rather qualitative. However such quantitative results allow to put
in a firm theoretical bases some of recent results related with recent
SFH inferences. So, instead to perform quantitative test, we use the
results by other authors to discuss our main results.

{\bf 1.} {\it Extending \cite{BBP14} results.}  The first result
refers to the age $\mathrm{t_{test}}$ that should be used to calibrate
recent SFH indices. As it has been shown, the best $\mathrm{t_{test}}$
value is the age of the galaxy under consideration $\mathrm{t_{age}}$
(which actually is redshift dependent). It applies even for SFR
inferences in regions inside galaxies, since it is always posible the
contribution of an old stellar population.

Taken that into consideration, we can extend the results obtained by
\cite{BBP14} about the use of any particular $\mathrm{t_{test}}$:
\cite{BBP14} used the SFH from MIRAGE simulations \citep{Peretal14}
covering ages up to 780 Myr and compare the instantaneous SFH with the
evolution of $\left<{\cal{SFR}}\right>_{\lambda}$ for different
indices ($Q(H)$, FUV, NUV and $u$) obtained by including the simulated
SFH in stellar population synthesis codes.  Their main finding is that
the calibration of the SFR is age dependent (i.e. in line with our
claim that the best $\mathrm{t_{test}}$ is the age of the system), and
they propose to use of a $\mathrm{t_{test}}$ of at least 1 Gyr instead
the typical one of 100 Myr when a fixed value of $\mathrm{t_{test}}$
is used. We note that 1 Gyr is nearby the maximum age considered by
their used SFH.

However, we can also establish that, extending the simulations over
larger age range, a calibration over $\mathrm{t_{test}} = 1
~\mathrm{Gyr}$ will produce again biased results (see
Sect.~\ref{sec:hat+pl}). In particular, the case of the $u$ band is
specially ill defined as SFR index: since it evolves as a power law
with slope close to the limiting value of $-1$, it would looks to be a
good SFR index for any fixed age $\mathrm{t_{test}}$, but it
overestimate the true SFR if the system is older than
$\mathrm{t_{test}}$.

In addition, \cite{BBP14}Ê~studied the delay between $\psi(t)$ and the
$\left<{\cal{SFR}}\right>_{\lambda}$ produced by the models at the
given $t$. They found that $\left<{\cal{SFR}}\right>_{Q(H)}$ follows
$\psi(t)$ with delay of around 1 Myr, whereas the other
$\left<{\cal{SFR}}\right>_{\lambda}$ indices have typical delays of
few Myr, although their plots (e.g. Figs. 6 and 8) shows that there is
a delay plus an smoothness effect. Such results are,again, fully
consistent with our analysis where
$\left<{\cal{SFR}}\right>_{\lambda}$ is a filter over $\psi(t)$.

{\bf 2.} {\it \cite{Johetal13} results.}  A second result is to break
the artificial duality in the use of $\mathrm{t_{test}}$, that is
implicitly assumed to be related with a possible value of
$\mathrm{t_{ind}}$, i.e. the time scale over the SFR is averaged. We
have shown that such time-scales can not be obtained, since it depends
on the particular SFH, which is unknown. Ever more, to impose {\it ad
  hoc} a constant SFH to obtain a $\mathrm{t_{ind}}$ value produces
ill-defined questions, since $\mathrm{t_{ind}}$ is intrinsically
undefined.  This situation is clearly illustrated in \cite{Johetal13}
who, making use of SFH obtained from CMD, computed the SFH dependent
$t_{uv,80\%,\psi(t)}$ values from a sample of 50 nearby dwarf galaxies
(where $uv$ refers to both FUV and NUV). They find that depending on
the SFH such values ranges from few Myr up to 10 Gyr, being this value
linearly correlated with the NUV-$r$ color, so the inferred
$\left<{\cal{SFR}}\right>_{uv}$ can not be univocally related with any
SFR time-scale.

We stress that such result is not a problem of the calibrations of the
recent SFR, but rather with the interpretation of what we would like a
$\left<{\cal{SFR}}\right>_{uv}$ value provides, but it does
not. Again, the calibrations are correct (when $\mathrm{t_{test}} =
\mathrm{t_{age}}$, a question also addressed partially in
\citealt{Johetal13}), but such calibrations does not provide directly
a time scale; it is required additional information (as optical or IR
colors) to provide a recent SFR time scale (actually information about
the global SFH). As an example, our computations produce a NUV-$r$ =
1.57 (c.f. Tab.~\ref{tab:AB}) with a characteristic age associated to
such color around 180 Myr. In previous section we stated that a bluer
(redder) NUV-$r$ indicates that the SFH is more concentrated at
younger (older) ages, which translate to a lower (larger) value of any
$t_{\mathrm{uv},x\%,\psi(t)}$ characteristic age; an effect which is
in agreement with \cite{Johetal13} findings.

However, we note that our explanation of the correlation between
NUV-$r$ and $t_{\mathrm{uv},80\%,\psi(t)}$ found by \cite{Johetal13}
is only valid for NUV-$r$ color bluer or near a NUV-$r$ value of 1.57,
but it cannot extended to extreme (much redder than 1.57) NUV-$r$
colors. That is, we only explain the bluer part of the correlation
found by \cite{Johetal13}, but it is required a more complete study
about the impact of the SFH at old ages (roughly, larger than 1 Gyr)
to find a satisfactory explanation of the correlation.

\begin{figure*}
\resizebox{\hsize}{!}{{\includegraphics{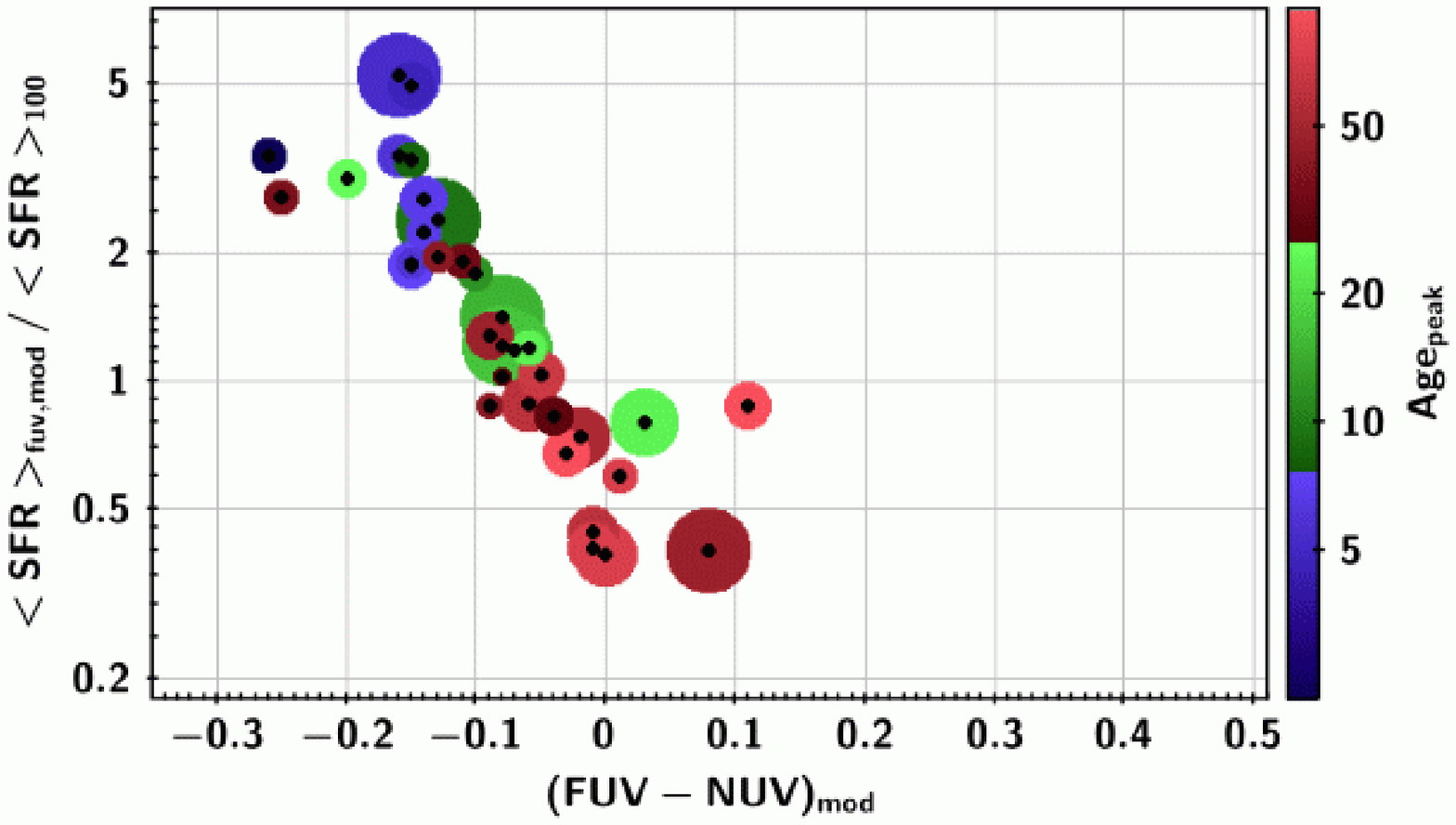}}{\includegraphics{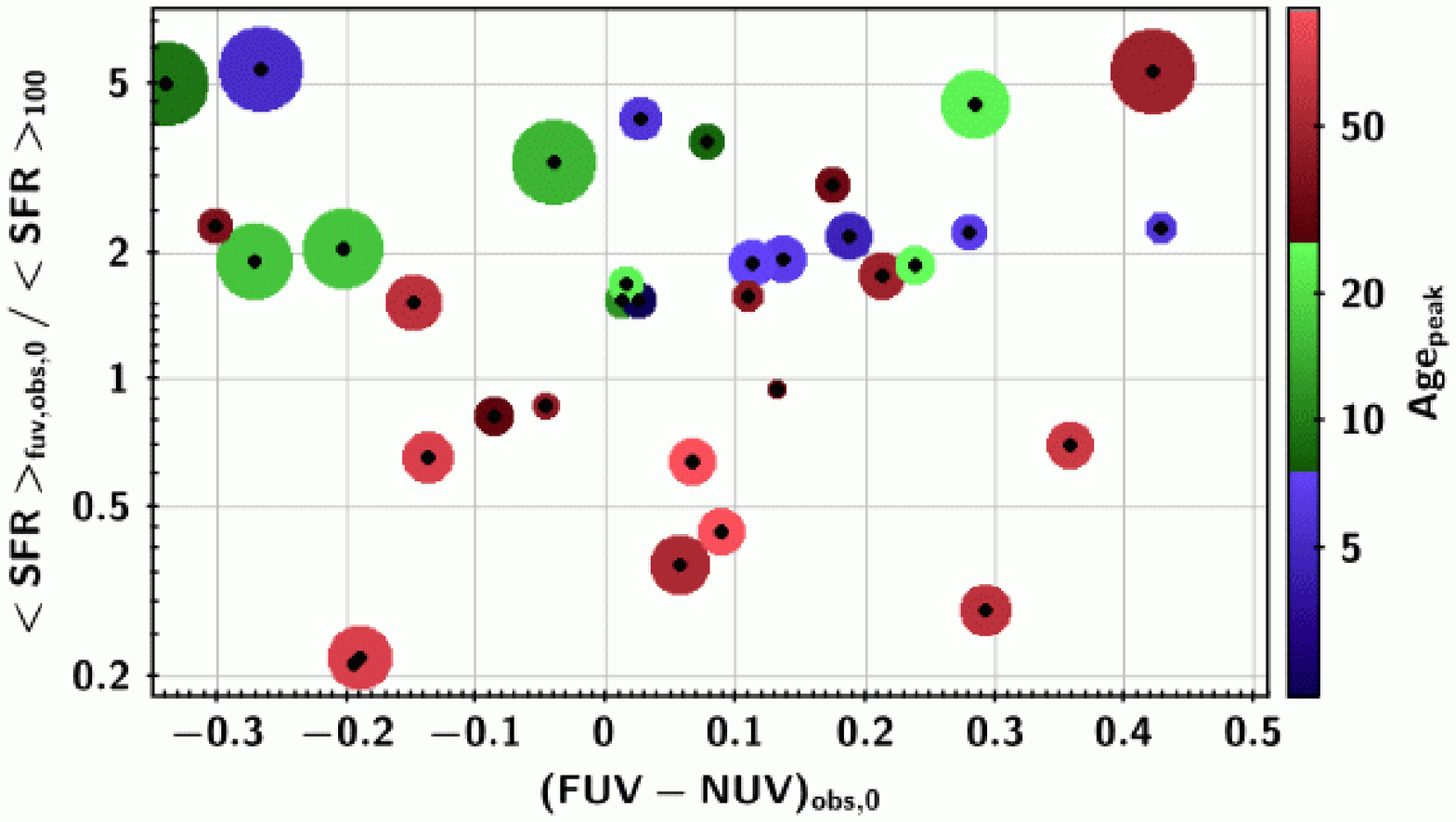}}}
\caption[]{Ratio of $\log \left<{\cal{SFR}}\right>_{fuv}/
  \left<SFR\right>_{100}$ vs. the FUV-NUV color obtained from
  \cite{Simetal14} data by the use of the SFH implemented in synthesis
  models (left), and obtained from the observed data once corrected
  from extinction (right). The color of the different points shows the
  Age$_\mathrm{peak}$ value, and the size of each point is
  proportional to $M_\mathrm{peak}/M_{100}$.}
\label{fig:setal14}
\end{figure*}

{\bf 3.}  {\it \cite{Simetal14} results.}  In the case of star forming
regions inside a galaxy we have a similar situation of a correlation
of different colors with any SFR averaged over a predefined
time-scale, although with some subtle differences: (a) Stellar
populations formed at old ages will be spread over all the volume of
the galaxy, hence, it is expected that $\psi(t)_\mathrm{region}$ that
would be obtained from particular blue region will have a lower
contribution from the older stellar populations (modulus the position
in the galaxy). (b) Although the increasing of resolution would
optimize a $\psi(t)_\mathrm{region}$ inference, it also implies a
reduction in the amount of stars which contributes to the total
luminosity, so an increasing on the uncertainty of the inferences
obtained from the integrated luminosity \cite[the so called IMF
  sampling effects, although stellar luminosity function sampling
  effects is a more correct description; see][ and references therein
  for a extensive discussion on the subject]{CL04,CLCL06,Cer13}.

Let us illustrate both situations using the work by \cite{Simetal14},
who analyzed the CMDs obtained from the Panchromatic Hubble Andromeda
Treasury data \citep{Daletal12} to obtain the corresponding SFH in the
last 500 Myr, and extinction of 33 FUV-bright regions in M31 and use
them to test the reliability of FUV as an SFR index at small scales.

The authors provides the SFH of each region; from it, they obtain the
SFH averaged over the last 100 Myr ($\left<SFR\right>_{100}$), the age
where the SFH has a peak, Age$_\mathrm{peak}$, and the ratio between
the mass of stars formed in the Age$_\mathrm{peak}$ over the mass of
star formed in the las 100 Myr, $M_\mathrm{peak}/M_{100}$.  In
addition, they use the SFH as input of a synthesis model to obtain the
integrated luminosity in {\sc galex}/FUV and (FUV-NUV)$_\mathrm{mod}$
color, and the corresponding
$\left<{\cal{SFR}}\right>_{fuv,\mathrm{mod}}$ using the standard
calibration. Finally, they use their extinction solution and apply it
to {\sc galex} data to obtain the extinction corrected FUV flux and
the corresponding $\left<{\cal{SFR}}\right>_{fuv,\mathrm{obs,0}}$.
One of the advantages of this paper is that, besides their detailed
analysis, the authors provide a plot the SFHs obtained from each of
the studied region as well as different set of tables including the
computed quantities, from which not tabulates values, as the
extinction corrected (FUV-NUV)$_\mathrm{obs,0}$ color, can be
obtained.  From a comparison of the ratio $\log
\left<{\cal{SFR}}\right>_{fuv}/ \left<SFR\right>_{100}$ as a function
of the area covered by the region, and using observed and modeled
$\left<{\cal{SFR}}\right>_{fuv}$ values, they claim that the
extinction corrected FUV fluxes are, on average, consistent with
$\left<SFR\right>_{100}$ within a 1-$\sigma$ scatter, which is related
with the discrete sampling of the IMF and the high time variability on
the recent SFH.

Again we can extend the conclusions of \cite{Simetal14} taking
advantage of the present study.  In Fig.~\ref{fig:setal14} we show the
ratio $\log \left<{\cal{SFR}}\right>_{fuv}/ \left<SFR\right>_{100}$
vs. the FUV-NUV color obtained from \cite{Simetal14} by the use of the
SFH implemented in synthesis models (left), and obtained from the
observed data once corrected from extinction (right). The color of the
different points shows the Age$_\mathrm{peak}$ value, and the size of
each point is proportional to $M_\mathrm{peak}/M_{100}$.

When synthesis models are used and sampling effects are neglected,
there is a clear correlation between $\log
\left<{\cal{SFR}}\right>_{fuv}/ \left<SFR\right>_{100}$, the FUV-NUV
color and Age$_\mathrm{peak}$, which is stronger for larger
$M_\mathrm{peak}/M_{100}$. The combination of Age$_\mathrm{peak}$ and
$M_\mathrm{peak}/M_{100}$ are a measure about the concentration of the
SFH at different ages, so the results of their simulations are
consistent our prediction about the dependence of
$\left<{\cal{SFR}}\right>_\lambda$, $\left<SFR\right>_{\Delta t}$, and
the color of the system. We note that the \cite{Simetal14} conclude
that the dispersion on $\left<{\cal{SFR}}\right>_{fuv}/
\left<SFR\right>_{100}$ are due to to the the variability of the
recent SFH, but they are not aware about the correlation shown here
and that such correlation can be used to reduce such scatter.

In the case of use observational data, sampling effects produce that
the correlation of $\log \left<{\cal{SFR}}\right>_{fuv}/
\left<SFR\right>_{100}$ and the FUV-NUV color disappear. This result
is not surprising since only 1 cluster in their analysis reach an
amount of gas transformed into stars in the last 100 Myr larger than
$10^{5}$ M$_\odot$, and such value is roughly the lowest limit quoted
by \cite{CL04} to model a system safely in UV-Optical bands
(i.e. without extreme sampling effects where the mean value obtained
by synthesis models lost its predictive power). However, there is
still a clear tendency of found lower values of
$\left<{\cal{SFR}}\right>_{fuv}/ \left<SFR\right>_{100}$ in clusters
where the SFH has a larger star formation concentration at older ages
and viceversa. It is, $\left<{\cal{SFR}}\right>_{fuv}/
\left<SFR\right>_{100}$ still depends on the age range where the
actual SFH is more concentrated.

\section{Conclusions}
\label{sec:conclu}

In this work we had translated the statements quoted in the constant
SFR approximation presented by \cite{Ken98}, which requires synthesis
models for its calibration, to the intrinsic algebra of synthesis
models in order to capture the principal characteristics of such
approximation which allows to obtain reasonable SFR inferences. The
results obtained from this study are:

\begin{enumerate}
\item When expressed in terms of SFH estudies, any integrated
  luminosity can be (and should be) considered as the result of
  filtering the SFH using SSP.

\begin{eqnarray}
\left<{\cal{SFR}}\right>_{\varphi_\lambda(t)} &=&
\int_0^{\mathrm{t_{age}}} \, \psi(t) \, \varphi_\lambda(t) dt =
\nonumber \\ &=& \frac{\int_0^{\mathrm{t_{age}}} \psi(t)\,
  \ell_{\lambda,\mathrm{IMF}}(\mathrm{t_{age}} - t)\,
  \mathrm{d}t}{{\int_0^{\mathrm{t_{age}}}
    \ell_{\lambda,\mathrm{IMF}}(t_*)\, \mathrm{d}t_*}} \nonumber \\ &
=&
\frac{{\cal{L}}_\lambda(\mathrm{t_{age}})}{\int_0^{\mathrm{t_{age}}}
  \ell_{\lambda,\mathrm{IMF}}(t_*)\, \mathrm{d}t_*} = C_{\lambda}
\times {\cal{L}}_\lambda(\mathrm{t_{age}}),
\label{eq:meanSFRfin}
\end{eqnarray}

\noindent being $C_\lambda$, the SFR calibration coefficient, a
normalization factor of SSP models.

\item Given that all the SFH of the system must be taken into account,
  the most reliable choice of the age to be used in the calibration is
  the system age, $\mathrm{t_{age}}$ (roughly 13 Gyr at $z=0$). This
  calibration varies with the redshift provided correct the assumption
  that all galaxies had been formed at a given cosmic epoch and
  independently of their posterior SFH.

\item The time evolution of the SSP luminosity
  $\ell_{\lambda,\mathrm{IMF}}(t)$ from 0 to $\mathrm{t_{age}}$ acts
  like a filter over the SFH, so it is the characterization of
  $\ell_{\lambda,\mathrm{IMF}}(t)$ who enables us to infer recent
  SFR. Under this perspective, there is no requirement about the
  functional form of the SFH to calibrate different SFR indices; in
  particular, a constant SFR is not a required hypothesis. The only
  advantage of the a constant SFH assumption is that, if Nature had
  works in such a way, the resulting SFR is an exact value.

\item Using a simple, parametrization of the SSP luminosity evolution
  $\ell_{\lambda,\mathrm{IMF}}(t)$, and detailed synthesis models
  results, we found that $U$ band is a ill-defined index to be used as
  a primary proxy of the SFR. It looks like primary proxies ($Q(H)$ or
  UV indices) when the calibration is done using small time scales,
  and as optical indices when used large time scales. Whatever the
  case such situation does not pose any problem if $\mathrm{t_{age}}$
  is used as calibration age.

\item We had shown that the assumed requirement that the integrated
  luminosity reach an asymptotical or steady-state value under a
  constant SFR hypothesis is not needed. Actually, for the given age
  of the Universe, such asymptotical value is never reached. Reach the
  asymptotical would allow to define a practical cut-off in the
  sensitivity defined by $\ell_{\lambda,\mathrm{IMF}}(t)$, hence to
  define a characteristic time scale over the SFH is in practice
  averaged. Unfortunately such cut-off does not exists and
  characteristics time scales are dependent on the unknown SFH. The
  best be can do is to characterize the sensitivity to the SFH
  provided by $\ell_{\lambda,\mathrm{IMF}}(t)$. We have shown that the
  time used for the calibration must be not confused with the
  characteristic time scales of $\ell_{\lambda,\mathrm{IMF}}(t)$ which
  are strongly dependent on the wavelength.  We have provide different
  ways to obtain such characteristic time-scales.

\item Using the $\left<{\cal{SFR}}\right>_{\lambda}$ values obtained
  from different indices and the characterization
  $\ell_{\lambda,\mathrm{IMF}}(t)$ (e.g. the use of equivalent widths
  or colors), we can establish time ranges where the SFH have a larger
  contribution to the different indices, hence improve the meaning of
  the measure given by $\left<{\cal{SFR}}\right>_{\lambda}$. The
  results obtained in this way are independent of the functional form
  of the SFH. To perform this task, it is required to calibrate all
  possible wavelengths (not only the standard ones of ionizing flux or
  UV fluxes) as established by Eq.~\ref{eq:meanSFRfin}.
 
 \item We have shown that, theoretically, there should be a
   correlation between the SFR obtained by the calibration of a
   particular luminosity $\left<{\cal{SFR}}\right>_{\lambda}$, the
   physical SFR which is the SFH averaged over a given time interval
   $\left<{\cal{SFR}}\right>_{\Delta t}$ and the galaxy colors. Such
   correlation are present in other works in the literature, and it is
   generally considered has a prove of the different time scales
   associated to $\left<{\cal{SFR}}\right>_{\lambda}$ and
   $\left<{\cal{SFR}}\right>_{\Delta t}$, hence a problem to obtain
   $\left<{\cal{SFR}}\right>_{\Delta t}$. We show that it is a natural
   result implicit in the very nature of the relation of the observed
   luminosity and the SFH of the system, and that it can be used to
   correct (or at least estimate a correction) of
   $\left<{\cal{SFR}}\right>_{\lambda}$ to obtain
   $\left<{\cal{SFR}}\right>_{\Delta t}$.

\end{enumerate}

After this study we conclude that the constant SFR approximation
quoted by \cite{Ken98} actually contains deeper implications which are
intrinsic to the population synthesis model algebra, but with a
different wording and a few subtle changes: (1) The quoted constant
SFH assumption is naturally translated to a normalization factor to
express SSP results as a sensitivity curve, and it is applicable to
any wavelength. (2) The steady-state (i.e. asymptotic) requirement to
define reliable SFR is naturally translated to a measure of the
relative sensitivity of the $\ell_{\lambda,\mathrm{IMF}}(t)$ filter to
the young and old component of the SFH, and, although a desirable
property, it is not a requirement to obtain information about the
recent SFH. (3) Finally, the bluest, the best statement is a synthetic
and operative version about the fact that, whatever the wavelength,
there is a peak of sensitivity in the recent SFH age range. Since
shorter wavelengths have a larger sensitivity, a blue color assures
that the possible contamination from the old component of the SFH is
minimized. However such statement has a limit depending on the galaxy
color and the studied system; it works for systems with colors redder
than the colors associated to the calibration of
$\ell_{\lambda,\mathrm{IMF}}(t)$ (or equivalently, predictions of a
constant SFH over all the possible age range). In the case extreme
blue colors, there is a first order correlation within the color, the
obtained value of $\left<{\cal{SFR}}\right>_{\lambda}$ and the actual
value of $\left<{\cal{SFR}}\right>_{\Delta t}$. It is not clear if
such correlation can be used to transform
$\left<{\cal{SFR}}\right>_{\lambda}$ value into the desired value of
$\left<{\cal{SFR}}\right>_{\Delta t}$, but at least provides hints
about over or underestimations of $\left<{\cal{SFR}}\right>_{\lambda}$
with respect $\left<{\cal{SFR}}\right>_{\Delta t}$.

As a final comment, this work has been done in an old-fashion way,
preferring the use of reasonable analytical approximations as a
function of suitable parameters to the use of detailed numerical
computations where numerical values difficult any possible
parametrization.  Such kind of reasoning, although not exact, can be
found in most of B. Tinsley papers, and A. Buzzoni ones who show that
the key points to understand the results obtained by detailed
simulations can be obtained using simple, but powerful, reasoning. As
we had shown, such methodology provide hints about which kind of plots
or correlations would be hidden under more elaborated numerical
experiments. It is true that for some aspects (track interpolations,
atmosphere models assignation, among others) synthesis models should
be used as black boxes for non initiated developers, but for some
purposes a simple inspection of the implicit equations in any
synthesis model, and their possible solutions, is the only
requirement.

\begin{acknowledgements}
We thank the referee, Rob Kennicutt, for his comments, which helped
improve this paper.  This work has done extensive use of TopCat
software \citep{topcat} and we acknowledge Mark Taylor for its
development.  This work has been supported by the Spanish Programa
Nacional de Astronom\'\i a y Astrof\'\i sica of the MINECO by the
projects AYA2014-58861-C3-1 (MC), and AYA2013-42781-P (SH), and
partially supported by the project AYA2011-C03-01 (AB).
\end{acknowledgements}

\end{document}